\documentclass[12pt]{article}
\usepackage{cite}
\usepackage{amsmath, amsthm, amssymb, graphicx,slashed}

\numberwithin{equation}{section}

\def\p{\partial}

\def\be{\begin{equation}}
\def\ee{\end{equation}}
\def\ba{\begin{align}}
\def\ea{\end{align}}
\def\beq{\begin{eqnarray}}
\def\eeq{\end{eqnarray}}
\def\a{\alpha}
\def\b{\beta}

\begin{document}

\begin{titlepage}

\vskip 1.5in
\begin{center}
{\bf\Large{Counting Strings, Wound and Bound} }\vskip 0.5cm
{ Sujay K. Ashok$^a$, Suresh Nampuri $^b$ and Jan Troost$^b$ } \vskip
0.3in

 \emph{$^{a}$Institute of Mathematical Sciences\\
   C.I.T Campus, Taramani\\
   Chennai, India 600113\\ }

 \vspace{.1in}

  \emph{\\${}^{b}$ Laboratoire de Physique Th\'eorique}\footnote{Unit\'e Mixte du CNRS et
     de l'Ecole Normale Sup\'erieure associ\'ee \`a l'universit\'e Pierre et
     Marie Curie 6, UMR
     8549.
 }\\
 \emph{ Ecole Normale Sup\'erieure  \\
 24 rue Lhomond \\ F--75231 Paris Cedex 05, France}
\end{center}
 \vskip 0.5in

\baselineskip 16pt

\begin{abstract}

  \vspace{.2in} We analyze zero mode counting problems for Dirac
  operators that find their origin in string theory backgrounds. A
  first class of quantum mechanical models for which we compute the
  number of ground states arises from a string winding an isometric
  direction in a geometry, taking into account its energy due to
  tension. Alternatively, the models arise from deforming marginal
  bound states of a string winding a circle, and moving in an
  orthogonal geometry.  After deformation, the number of bound states
  is again counted by the zero modes of a Dirac operator. We count
  these bound states in even dimensional asymptotically linear dilaton backgrounds 
as well as in Euclidean Taub-NUT. 
  We show multiple pole behavior in the fugacities 
 keeping track of a $U(1)$ charge. 
 We also discuss a second class of counting problems that arises when these backgrounds are deformed via 
 the application of a heterotic duality  transformation.
 We discuss applications of our results to Appell-Lerch sums and the counting of domain wall bound states. 
\end{abstract}
\end{titlepage}
\vfill\eject

\tableofcontents

\section{Introduction}
It is often useful to reduce a physical problem in a higher
dimensional theory to a property of a lower dimensional model that is
more easily solved. 
In this paper, we concentrate on the calculation of 
Dirac indices in a class of supersymmetric quantum mechanics models. These
supersymmetric quantum mechanics problems arise when we study the
motion of a string on a target manifold with an isometry. When we wind
the string in the isometric direction, we perform a reduction from $1+1$ dimensions
to $1$ dimension, and we obtain a supersymmetric
quantum mechanics with potential
\cite{AlvarezGaume:1983ab}. The ground states
correspond to 
bound states of wound strings. 

Another context in which the same
model arises is when we study the non-trivial space with isometry,
tensored with a circle. When we now wind the string on the circle, it
is often marginally bound to the non-trivial space, while preserving
supersymmetry. When reducing our attention to its center of mass
motion on the geometry, it gives rise to a supersymmetric quantum
mechanics. It is then interesting to find a natural deformation of the
resulting supersymmetric quantum mechanics such that the marginal
bound state spectrum is deformed to a spectrum of bound states. We can
then proceed to count the latter. The natural deformation consists
in introducing a small winding along the isometric direction, as
above. 

We employ also a second deformation technique 
in which we deform the background itself. We imagine embedding our solution in some
heterotic or Kaluza-Klein reduced supergravity theory and implement a
small duality transformation on the isometric direction and some gauge
direction in the supergravity theory, to introduce a non-trivial gauge
field in the supersymmetric quantum mechanics.  We then consider
charged excitations, coupled to the gauge field.  Both techniques that
we describe enter in the realm of the general technique of adding
potentials to moduli spaces.

We apply our two deformation techniques to a class of string
backgrounds which are non-compact and which have an asymptotically
linear dilaton. They were first written down in \cite{Kiritsis:1993pb}
and were later studied in the context of gauged linear sigma models
\cite{HK1, HK2}. We will refer to these as asymptotically linear
dilaton  spaces. The simplest example of such a space is the well
studied supersymmetric cigar conformal field theory in two
dimensions. Index theorems for these non-compact backgrounds need to
take into account subtle boundary contributions.  We find it
convenient to proceed by an explicit counting of states, and as a
bonus provide a
 construction of the bound state solutions in
these backgrounds.

We also apply the methods we develop to the Euclidean Taub-NUT space
in four dimensions, and make a technical improvement to the existing
method \cite{Pope} of counting bound states. The latter index has
already found applications to the counting of monopole bound
states \cite{Gauntlett:1999vc} and to the counting of one quarter BPS
dyons in ${\cal N}=4$ four-dimensional compactifications of string
theory \cite{David:2006yn}. Likewise, we will highlight 
potential applications of our results on asymptotically linear dilaton backgrounds in the realm of domain wall bound states in gauge theories.

Another application of the index formulas
is a re-derivation of the mock modular form that arises in the
calculation of the elliptic genus of the cigar conformal field
theory. This is done by first defining a generating function that
keeps track of the Dirac zero modes in a given momentum and winding
sector. This generating function is then lifted to the elliptic
genus by assigning appropriate 
fugacities that keep
track of the momentum and winding quantum numbers, and by adding oscillator
contributions. The index calculation gives a direct physical understanding
of the double sum form of Appell-Lerch sums featuring in physical counting
problems.

Our paper is structured as follows. We start in section \ref{sqm} by
reviewing supersymmetric quantum mechanics with a potential on target
manifolds with an isometry
and set up the wound bound state
counting problem. In section \ref{lindil} we introduce the
asymptotically linear dilaton spaces of our interest and introduce all
the technology needed to compute the Dirac indices that depend on a winding
number and a momentum charge. 
In section \ref{het} we introduce the heterotic
backgrounds that we generate by duality transformations, and the
details on how to compute the indices in these
backgrounds. In the final section \ref{con} we conclude and indicate
possible applications of these counting formulas. Some technical details are
provided in appendices \ref{lineareqns} and \ref{Hopf}. In appendix \ref{ETN} we illustrate how
our techniques simplify the calculation of the index 
on Euclidean Taub-NUT with self-dual gauge field.

\section{Indices, quantum mechanics and field theory}
\label{sqm}
In this section, we review a class of supersymmetric quantum mechanics
models which are one-dimensional non-linear sigma-models. 
We recall the existence of a potential that is consistent with
supersymmetry when the target manifold of the model has an isometry, and how that
potential arises naturally from a twisted dimensional reduction from $1+1$
dimensions \cite{AlvarezGaume:1983ab}. The model then corresponds to the
center of mass motion of a wound string, and the potential arises from the
energy of the tensionful string. Secondly, we mention how a
marginally bound string, wound transversely to such a target space
geometry after a small deformation 
 also gives rise to the supersymmetric
quantum mechanics with Killing vector potential.

\subsection{The supersymmetric quantum mechanics : manifolds with isometry}
We consider a $0+1$ dimensional sigma-model with 
at least 
one supersymmetry, namely a supersymmetric quantum
mechanics. The target manifold $M$ with metric $G$ permits
one Killing isometry generated by the vector field $K$. We have
therefore the Killing equation:
\be
{\cal L}_K G = 0.
\ee
The Lagrangian of the model includes a kinetic term and a potential term associated
to the existence of the Killing isometry \cite{AlvarezGaume:1983ab}:
\be\label{SQMwithK}
L = \frac{1}{2} (G_{\mu \nu} \dot{X}^\mu \dot{X}^\nu +
i G_{\mu \nu} \psi^\mu {\cal D}_\tau \psi^\nu - G_{\mu \nu} K^\mu K^\nu
- i D_\mu K_\nu \psi^\mu \psi^\nu).
\ee
The dot represents a derivative with respect to the world line time $\tau$,
while the derivative ${\cal D}_\tau$ is the covariant derivative of the tangent
space valued world line fermion.
If we define the vielbein $e^a_\mu$, and the world line fermion $\psi^a = \psi^\mu e_\mu^a$,
then 
after quantization we find the canonical commutators:
\begin{eqnarray}
{[} X^\mu , p_\nu {]} &=&  i \delta^\mu_\nu
\nonumber \\
\{ \psi^a , \psi^b \} &=& \delta^{ab}.
\end{eqnarray}
The fermionic modes form a Clifford algebra. After quantization, they give rise to a Hilbert space
corresponding to a space-time fermion.
The super-covariant momentum:
\be
\pi_\mu = p_\mu - \frac{i}{4} \omega_{\mu ab} {[} \psi^a , \psi^b {]}
\ee
acts like a covariant derivative on space-time spinors: $\pi_\mu = -i D_\mu$.
The supercharge $Q$ is given by the formula:
\be
Q = \psi^\mu (-i D_\mu - K_\mu),
\ee
while the central charge $Z$ is:
\be
Z = K^\mu \pi_\mu - \frac{i}{2} ( D_\mu K_\nu) \psi^\mu \psi^\nu.
\ee
It is proportional to the Lie derivative acting on spinors. The
supersymmetric quantum mechanics allows for a $\mathbb{Z}_2$ grading
(world line fermion number) given by the
operator $\prod_a \psi^a$ which anti-commutes with the Dirac operator:
\be
\slashed{D} = \sqrt{2} Q = \gamma^\mu ( - i D_\mu -K_\mu).
\ee
In most of our paper, we concentrate on computing the ground
state contributions to the Witten index, namely the difference between
the number of world line bosonic and fermionic zero modes of the Dirac
operator in a given set of models.  
The index is defined for each fixed central charge sector.

\subsection{Twisted dimensional reduction}
In this section, we recall how the above supersymmetric quantum mechanics model can be obtained
from a Scherk-Schwarz reduction \cite{AlvarezGaume:1983ab}. 
We start with a $(1,0)$ supersymmetric theory in $1+1$ dimensions. We will consider
a compact
space-like $\sigma$-direction of period $2 \pi$.
 We take the action to be:
\be
S = \frac{1}{4 \pi} \int d^2 \sigma (G_{\mu \nu} ( \frac{1}{\alpha'}
(\partial_\tau X^\mu \partial_\tau X^\nu - \partial_\sigma X^\mu \partial_\sigma X^\nu)
+ i \psi^\mu { \cal D}_{\tau - \sigma} \psi^\nu )). \label{oneplusone}
\ee
Note that the choice of sign of derivative with respect to $\sigma$ in
the fermion kinetic term determines whether we concentrate on the
supersymmetric quantum mechanics associated to the $1+1$ dimensional right-
or left-movers. %

Let us consider the target space isometry generated by the vector field
$K$. Scherk-Schwarz or twisted dimensional reduction can be carried out by
requiring that the dependence of the fields on the 
$\sigma$ direction is
given in terms of the Killing vector that generates the isometry:
\be
\p_{\sigma} X^{\mu} = 
K^{\mu} \qquad \text{and}\qquad \p_{\sigma}\psi^{\mu} = -\p_{\nu}K^{\mu}\psi^{\nu}  \,.
\ee
We now integrate over the $\sigma$ direction and obtain a quantum
mechanical system. Our ansatz ensures that the reduction will
be supersymmetric. Plugging this ansatz into the action, we find the
dimensionally reduced action:
\be
S =  \frac{1}{2} \int d \tau (G_{\mu \nu} \frac{1}{\alpha'}
\partial_\tau X^\mu \partial_\tau X^\nu 
+ i G_{\mu \nu} \psi^\mu { \cal D}_{\tau} \psi^\nu ))
- \frac{G_{\mu \nu}}{\alpha'} K^\mu K^\nu
+ i \psi^\mu K_{\mu;\nu} \psi^\nu\,,
\ee
which is the action in equation \eqref{SQMwithK} provided we put
$\alpha'=1$, which we will do from now on.
The Scherk-Schwarz reduced theory has the interpretation of
corresponding to the center of mass motion of a string wound along the
isometric direction. This is a natural way in which the
supersymmetric quantum mechanics problem that we study arises in
string theory. 

\subsection{A marginal bound state problem}
As an aside, we want to mention a second way in which the same
supersymmetric quantum mechanics problem pops up in string theory.  We
consider a string living in the space-time $\mathbb{R} \times M \times
S^1 \times X$ and wound on the $S^1$ circle. In many instances, a
fundamental or D-string wound on the circle is marginally bound to the
geometry $M$. If we suppose that the geometry $M$ admits a $U(1)$
isometry with compact action, we can associate to it a circle
$\tilde{S}^1$.  It is natural to study the (often truly) bound state
counting problem that arises from mixing the $S^1$ and $\tilde{S}^1 $
circles. That introduces a
potential on the manifold $M$, arising from the winding energy of the
string along the $\tilde{S}^1$ circle. If the size of the
$\tilde{S}^1$ circle shrinks towards the center of the geometry, the
string will now be bound to the center of the geometry $M$. 

We can put this idea into practice by starting out with a metric on $M
\times S^1$:
\be
ds^2 = G_{MN}dX^M dX^N = G_{\mu \nu} d X^\mu d X^\nu + R^2 dy^2.
\ee
The coordinate on the $S^1$ circle of radius $R$ is $y \equiv y + 2 \pi$. 
By assumption, the metric $G_{\mu \nu}$ on the manifold $M$ admits a Killing isometry $K$. Clearly
there is another Killing vector, namely $L=w_y  \p_y$ where $w_y$ represents the winding
number of the string around the circle $S^1$. We now repeat the
Scherk-Schwarz reduction on the worldsheet with the
$\sigma$-dependence given by
\be
\p_{\sigma}X^M = \sin\a K^M + \cos\alpha L^M \,. 
\ee
After neglecting constant terms, and allowing for no excitations in the decoupled $y$-direction,
we again wind up with the world line supercharge:
\be
Q= G_{\mu\nu} \psi^{\mu}(-i D_\mu - \sin \a K^{\mu}).
\ee
As before the index counting the number of bosonic minus fermionic
ground states is the index of the equivariant Dirac operator
$\slashed{D}_K = \slashed{D}- i \sin \a\slashed{K}$. We can
alternatively interpret the (dual of the) Killing vector field $A_\mu
= \sin \alpha \, G_{\mu \nu} K^\nu$ as an abelian gauge field on the
manifold $M$.

Our discussion in this section was very general, and can be viewed as
one more way to introduce potentials in supersymmetric quantum
mechanics, an ubiquitous tool. The technique we introduced above is
natural in many string theory contexts.  For an example application of
this tool with Euclidean Taub-NUT as the target space manifold, see
\cite{David:2006yn}.  In the following, we evaluate the index for new
example geometries $M$ with Killing vector fields $K$. We will compute
the index through 
the
explicit solution of the massless Dirac equation on
the target manifold $M$.

\section{Asymptotically linear dilaton spaces}
\label{lindil}
In this section, we introduce the target
spaces $M$ on which we concentrate. As string backgrounds, these backgrounds come accompanied by 
an asymptotically linear dilaton. The manifolds exhibit at least a $U(1)$ isometry
group. These 
supergravity solutions have been found in \cite{Kiritsis:1993pb}  and
discussed further in \cite{HK1, HK2}. 
Together with an internal conformal field theory,
 these backgrounds admit microscopic string theory interpretations as
near horizon geometries T-dual to configurations of
NS5-branes \cite{HK2}. The two-dimensional example is the cigar geometry
\cite{Elitzur:1991cb, Mandal:1991tz, Witten:1991yr}. We will later
show that at least 
in this example, the index calculation that we perform has an interesting two-dimensional
conformal field theory application.

\subsubsection*{The backgrounds}
The metric and dilaton of the supergravity solutions have the form
\begin{align}\label{dmetric}
ds^2 &= \frac{g_N(Y)}{2}dY^2 + \frac{2}{N^2 g_N(Y)}(d\psi + NA_{FS})^2 + 2Y ds^2_{FS}   \cr
\Phi &= -\frac{NY}{k} \,.
\end{align}
The function $g_N(Y)$ is given by: 
\be\label{generalgN}
g_N(Y) = \frac{Y^{N-1}}{N}\frac{e^{\frac{2NY}{k}}}{\int_0^Y t^{N-1} e^{\frac{2Nt}{k}} dt} \,.
\ee
The metric $ds^2_{FS}$ refers to the Fubini-Study metric on the complex projective space
$\mathbb{CP}^{N-1}$ and the connection one-form
$A_{FS}$ has differential equal to the K\"ahler form.
The backgrounds we discuss therefore have a $SU(n) \times U(1)$
isometry. The $U(1)$ factor refers to translations along the $\psi$ direction
and this is the Killing isometry $K$ we use to twist the Dirac operator. The
one-form gauge field $A$ dual to the Killing vector takes the form:
\be\label{oneformgeneraln}
A = \frac{2 w}{N g_N(Y)}(d\psi + N A_{FS}) \,,
\ee
where $w$ is the number of times our string winds the circle parametrized
by the coordinate $\psi$ of period $2 \pi N$.
In the following, we concentrate on the cases $N=1,2,3$, which
correspond to target spaces $M$ of dimension two, four and six.  The
parameter $k$ is a free parameter in gravity, but is typically
quantized in string theory. After T-duality, it can be related to the
number of $NS5$ branes that generate the background \cite{HK1, HK2}.

\subsection{The two-dimensional cigar}

The first example we concentrate on is the two-dimensional cigar
background. This corresponds to a two dimensional conformal field
theory with non-trivial metric and dilaton. It has a description as a
gauged Wess-Zumino-Witten model, which is an exact conformal field
theory. The metric and dilaton read:
\begin{align}
ds^2 &= k %\alpha'
(d\rho^2 + \tanh^2\rho d\psi^2)\cr
\Phi &= - \log\cosh\rho \,.
\end{align}
It is equivalent to the case $N=1$ in our general discussion by the
coordinate transformation $\cosh \rho = e^{ \frac{Y}{k}}$. The
function $g_1(Y)$ is given by
\be
g_1(Y) = \frac{2}{k}\frac{e^{\frac{2Y}{k}}}{e^{\frac{2Y}{k}}-1} \,.
\ee
The gauge field dual to the angular Killing vector is: 
\be
A =\frac{2w}{g_1(Y)}\, d\psi \,. 
\ee

\subsubsection*{The massless Dirac equation}
Given these prerequisites, we  turn to the solution of the massless
 Dirac equation for a charged Dirac fermion $\Psi$ on our curved background:
\be
\gamma^{\mu}(\p_{\mu}+\frac{1}{4}\omega_{\mu}^{ab}\gamma_{ab}-i A_{\mu})\Psi = 0\,.
\ee
In order to solve the Dirac equation we choose the zweibein:
\be
e^1 = \sqrt{\frac{g_1(Y)}{2}} dY
\qquad 
e^2 = \sqrt{\frac{2}{g_1(Y)}} d \psi
 \,.
\ee
Our conventions for spinors in various dimensions are given in
Appendix \ref{spinors}. We work in the Weyl basis:
\be
\Psi = \begin{pmatrix}
\Psi_-\\
\Psi_+
\end{pmatrix} = e^{i n \psi}\begin{pmatrix}
G_1(Y) \\
G_2(Y)
\end{pmatrix}\,.
\ee
The solutions to the Dirac equation are then given by
\be
\Psi =
e^{i n \psi}\begin{pmatrix}
e^{\frac{Y}{k}(n-kw)}\, (g_1(Y))^{\frac{1}{4}-\frac{n}{2}} \cr
e^{-\frac{Y}{k}(n-kw)}\, (g_1(Y))^{\frac{1}{4}+\frac{n}{2}}
\end{pmatrix}\,.
\ee
To determine the allowed zero modes, we first study the behavior of
the wave functions near the tip and near infinity. Near the tip at $Y=0$, the
wave functions take the form
\be
\Psi \sim e^{i n \psi} \begin{pmatrix}
Y^{-\frac{1}{4}+\frac{n}{2}} \cr
Y^{-\frac{1}{4}-\frac{n}{2}}
\end{pmatrix}\,,
\ee
while asymptotically, we have the behavior
\be
\Psi \sim e^{i n \psi} \begin{pmatrix}
e^{\frac{Y}{k}(n-kw)} \cr
e^{-\frac{Y}{k}(n-kw)}
\end{pmatrix}\,.
\ee
We need to impose regularity at the tip and require normalizability at
infinity. That leads to the following constraints on the variables $n$
and $w$:
\begin{align}
\frac{1}{2} & \le n < kw  \quad \text{for negative chirality modes} \cr
kw &< n \le - \frac{1}{2}  \quad  \text{for positive chirality modes} 
\end{align}
Since we have a positive level $k>0$, this implies that the winding $w$ is positive
for
negative chirality modes while the winding is negative for positive
chirality modes.
Note that we can think of the asymptotic charge $n-kw$ as an imaginary radial momentum.
When $kw$ is half-integer (as is the momentum $n$), we can have zero radial momentum.
It is clear that if we slightly shift it, we should consider either the positive chirality mode
or the negative chirality mode as normalizable. Let's choose a regularization such that for negative
momentum $n$, we allow $n=kw$.

These results can be encoded in a partition sum $Z_2$ keeping track of the
positive and negative chirality spinorial zero-modes, weighted with
their momentum $n$ and winding $w$: 
\begin{align}
Z_2 &=
\left( \sum_{w<0} \sum_{n=[[kw]]}^{-\frac{1}{2}} - \sum_{w>0} \sum_{n=\frac{1}{2},\frac{3}{2},\dots}^{[[kw - \epsilon]]} \right )y_1^n y_2^w ,
\end{align}
where ${[[} x {]]}$ is the strict half-integer (i.e. element of $ \mathbb{Z} + 1/2$) smaller than or equal to
$x$ and $\epsilon$ is a small positive regulator.  We have introduced fugacities
$y_1$ and $y_2$ to keep track of the supersymmetric quantum mechanics labeled by the winding $w$, and the
conserved charge $n$. The formal sum is convergent for particular values
of $y_1$ and $y_2$.
One can rewrite this double summation by first summing over the
momentum variable $n$. This leads to the final expression for the generating
function of indices:
\be\label{cigarZ2}
Z_2 = \left[\sum_{n < 0}\sum_{n-kw \ge 0} - \sum_{n>0}\sum_{n-kw <0}\right] y_1^n y_2^w \,.
\ee

Note that our approach to finding wound bound state solutions in the
cigar is complementary to the one of \cite{Dijkgraaf:1991ba} where
these bound states were uncovered by studying momentum modes in the
T-dual geometry. Here we use the Scherk-Schwarz reduction technique to
find  bound states. Our technique is more generally applicable
since it provides the means to compute bound states with both winding and
momentum. As we have argued in section \ref{sqm},
 it also generalizes to a large class of backgrounds.
\subsection{The four-dimensional asymptotically linear dilaton background}
The next example we tackle is the asymptotically linear dilaton solution in four dimensions:
\be\label{d=4metric}
ds^2 = \frac{g_2(Y)}{2} dY^2 + \frac{1}{2g_2(Y)}(d\psi + \cos\theta d\phi)^2 + \frac{Y}{2}(d\theta^2+ \sin^2\theta d\phi^2) \,.
\ee
The radial function $g_2(Y)$ is given by:
\be
g_2(Y) = \frac{8}{k^2} \frac{Y}{e^{-\frac{4Y}{k}}-1+\frac{4Y}{k} }\,.
\ee
Along with a non-trivial dilaton $\Phi=-\frac{2Y}{k}$, the background
solves the string equations of motion. The central charge of the
corresponding conformal field theory can be calculated from the dilaton profile and comes
out to be
\be
c = 6\left(1+\frac{4}{k}\right) \,.
\ee
The wave functions have a simple
dependence on the angular variables  $\psi$ and $\phi$. Given the form of the metric in
\eqref{d=4metric} the four Dirac components are separable and we choose
the ansatz
\be\label{Diracspinor4d}
\Psi = e^{in \psi +i m \phi}\begin{pmatrix}
G_1(Y) S_1(\theta)\\ 
\vdots \\ 
G_4(Y) S_4(\theta)
\end{pmatrix}
\ee
The explicit form of the $\gamma$-matrix basis
as well as the form of the Weyl spinors we will use has been given in
Appendix \ref{spinors}. We will solve the Dirac equation in terms of  the Weyl spinors $\Psi
= (\Psi_-, \Psi_+)^T$.

\subsubsection{Linear equations}

Since the fermion is massless we can consider the positive and
negative chirality spinors separately.  
To evaluate the spin connection we choose the vierbein:
\begin{align}
e^1 &= \sqrt{\frac{g_2(Y)}{2}} dY \qquad e^2 = \frac{1}{\sqrt{2g_2(Y)}}(d\psi + \cos\theta d\phi) \cr
e^3 &= \sqrt{\frac{Y}{2}} d\theta \qquad e^4 = \sqrt{\frac{Y}{2}} \sin\theta d\phi \,.
\end{align}
Substituting our ansatz for the positive chirality wave function
$\Psi_+ = e^{in \psi+im\phi}\bigl( \begin{smallmatrix}
  G_3(Y)S_3(\theta)\cr G_4(Y)S_4(\theta) \end{smallmatrix}\bigr)$ we
find that, apart from constant factors, the Dirac equation takes the
schematic form
\begin{align}
S_3(\theta)(L_3(x) \cdot G_3(x)) + G_4(x)(N_3(\theta)\cdot S_4(\theta)) &= 0 \cr
S_4(\theta)(L_4(x) \cdot G_4(x)) + G_3(x)(N_4(\theta)\cdot S_3(\theta)) &= 0\,.
\end{align}
There are similar equations for the negative chirality spinors. The
differential operators $L_i$ and $N_i$, in this case are given by
\begin{align}
-L_3 &= \sqrt{\frac{Y}{g_2(Y)}}\left(\frac{d}{dY}-n g_2(Y) + w \right)+ \left(\frac{g_2(Y)-Yg'_2(Y)}{4\sqrt{Yg_2^3(Y)}} \right)\cr
L_4 &= \sqrt{\frac{Y}{g_2(Y)}}\left(\frac{d}{dY}+n g_2(Y) - w \right)+ \left(\frac{g_2(Y)-Yg'_2(Y)}{4\sqrt{Yg_2^3(Y)}} \right)\cr
N_3&=\frac{d}{d\theta}-m\csc\theta+\left(\frac{1}{2}+n\right)\cot\theta\cr
N_4&=\frac{d}{d\theta}+m\csc\theta+\left(\frac{1}{2}-n\right)\cot\theta\,.
\end{align}
These equations are equivalent to second order equations for the radial and angular functions. As argued
 in appendix \ref{lineareqns}, it is sufficient to focus on the solutions to the linear set of equations
\be
L_3(Y)\cdot G_3(Y) = 0 =  L_4(Y)\cdot G_4(Y)\qquad\text{and}\qquad N_3(\theta)\cdot S_{4}(\theta)=0 = N_4(\theta)\cdot S_{3}(\theta) \,.
\ee
These have the following solutions:
\begin{align}
\Psi_+ &= e^{in \psi+im\phi}\begin{pmatrix}
C_3\, e^{-wY+\frac{2Yn}{k}}\left(\frac{8Y}{kg_2(Y)}\right)^{-\frac{1}{4}+\frac{n}{2}}
(\sin\theta)^{-\frac{1}{2}+n}(\tan\frac{\theta}{2})^{-m}\cr
C_4\, e^{wY-\frac{2Yn}{k}}\left(\frac{8Y}{kg_2(Y)}\right)^{-\frac{1}{4}-\frac{n}{2}}
(\sin\theta)^{-\frac{1}{2}-n}(\tan\frac{\theta}{2})^{m}
\end{pmatrix}
\end{align}

\subsubsection{The counting of zero modes}
\label{KKL4counting}
We must now identify the normalizable solutions.
The measure factor $\sqrt{g}$ is proportional to $Y \sin \theta$. Near
the tip at $Y =0$, we have that the wave-functions behave like $Y^{-\frac{1}{2}
  \pm n}$. The measure factor cancels the $Y^{-\frac{1}{2}}$ prefactor when we
square a wave-function component. A similar
phenomenon occurs for the angular factors. 

We also find that for positive momentum $n$,  the component $C_4$ needs to be zero
in order to have regular wave-functions.
Rewriting the
third component of the wave function in terms of the argument $\frac{\theta}{2}$, we obtain
\be
\Psi_{+,3} \sim (\sin\frac{\theta}{2})^{-\frac{1}{2}+n-m}\,  (\cos\frac{\theta}{2})^{-\frac{1}{2}+n+m}  \,.
\ee
The exponent of both factors needs to be positive, up to a shift by $-1/2$. We conclude that
$n \ge |m|$.

We moreover have that the  parity of $2n$ and of $2m$ needs to be opposite.
This is because near the tip of the space (which is locally a Euclidean four-plane)
the angles $\psi$ and $\phi$ are related to angles $\xi_{1,2}$
in two two-planes through the formulas $\psi=-\xi_1-\xi_2$ and
$\phi=\xi_2-\xi_1$. A $2 \pi$ rotation in one of these two-planes
must give a minus sign to the fermion wave-function.
All these constraints taken together
imply that the quantum number $m$ lies in the window
\be
-n+1/2 \le m \le n-1/2 \, ,
\ee
and jumps by integers when we fix $n$. 
For a given momentum  $n$, there is therefore a
degeneracy of $2n$ arising from the two-sphere. This
is the spin-degeneracy that arises from spherical harmonics with
spin $l=n-1/2$.

In order to understand the further constraints that arise from the radial part
of the problem, let us recall here the radial behavior of the function
$g_2(Y)$:
\be
Y^2\quad\xleftarrow{Y\rightarrow 0} \frac{Y}{g_2(Y)} \xrightarrow{Y\rightarrow \infty}\quad \frac{kY}{2}  \,.
\ee
For negative winding $w$ there are no normalizable solutions. For positive winding, the
wave functions have the following asymptotic behavior:
\be
 Y^{-\frac{1}{4}+\frac{n}{2}}\xleftarrow{Y\rightarrow 0}  \Psi_{+,3}\xrightarrow{Y\rightarrow \infty} e^{-Y\left(w-\frac{2n}{k}\right)} Y^{-\frac{1}{4}+\frac{n}{2}} \, .
\ee
Therefore imposing regularity and normalizability requires the momentum to be positive and in the range 
\be
\frac{1}{2} \le n\le \frac{kw}{2}
\ee
Note that we again regularize $n-kw/2$
 such that the extreme
case does contribute.

For negative momentum $n<0$, we find the spin degeneracy $2|n|$ and
negative winding. Only the component $\Psi_{+,4}$ will
be non-vanishing.
We need to impose the inequality:
\be
 \frac{kw}{2} <  n \le -\frac{1}{2} \,.
\ee
For negative chirality the zero mode solutions to the linear differential equations are given by
\be
\Psi_- = e^{in \psi+im\phi}\begin{pmatrix}
C_1\, e^{-wY+\frac{2Yn}{k}}\left(\frac{8Y}{kg_2(Y)}\right)^{-\frac{1}{4}+\frac{n}{2}}
(\sin\theta)^{-\frac{1}{2}-n}(\tan\frac{\theta}{2})^{m}\cr
C_2\, e^{wY-\frac{2Yn}{k}}\left(\frac{8Y}{kg_2(Y)}\right)^{-\frac{1}{4}-\frac{n}{2}}
(\sin\theta)^{-\frac{1}{2}+n}(\tan\frac{\theta}{2})^{-m} \, .
\end{pmatrix}
\ee
Noting the flip of signs in the angular wave functions relative to the
positive chirality wave functions, one can check that there are no
zero modes of negative chirality.

Finally, the partition sum that counts the zero modes, keeping track of
the momentum and winding quantum numbers, is given by
\begin{eqnarray} 
Z_4 &=& \left(\sum_{w>0}\sum_{2 n=1,2,\dots}^{{ [} kw {]}}
  +\sum_{w<0}\sum_{2n={[} kw-\epsilon{]}}^{-\frac{1}{2}}\right)\, 2
|n|\, y_1^{n}y_2^w
\nonumber \\
&=&
\left(\sum_{w>0}\sum_{2 n=1,2,\dots}^{{ [} kw {]}}
  -\sum_{w<0}\sum_{2n={[} kw-\epsilon{]}}^{-\frac{1}{2}}\right)\, 2
n\, y_1^{n}y_2^w \,, \end{eqnarray}
 where the double square brackets $[ x ]$
indicate the smallest integer smaller or equal to $x$.  The final
result is very similar to the index counting in Euclidean
Taub-NUT. Indeed, they have in common the transverse geometry
responsible for the degeneracy factor, as well as the tip region.  The
asymptotics of the spaces is different, yet they lead to the same
index. This aspect
 is discussed in Appendices \ref{lineareqns} and
\ref{ETN}.

\subsection{The six-dimensional asymptotically linear dilaton background }
We turn to the explicit construction of Dirac zero modes in the six-dimensional asymptotic
linear dilaton background (i.e. the case $N=3$ of background (\ref{dmetric})).

\subsubsection*{The background}

The metric and dilaton are given by
\begin{align}
ds^2 &= \frac{g_3(Y)}{2}dY^2 + \frac{2}{9 g_3(Y)}(d\psi + 3A_{FS})^2 + 2Yds^2_{\mathbb{CP}^2}\cr
\Phi &= -\frac{3Y}{k} \,.
\end{align}
The connection one-form $A_{FS}$ on $\mathbb{CP}^2$ has differential
equal to the Fubini-Study curvature two-form. In Appendix \ref{Hopf}
we give a detailed description of the geometry of $\mathbb{CP}^2$ as
well as a 
choice of coordinates and one-forms $e^{1,2,3,4}$.
Using these, the vierbein for the six-dimensional problem are chosen to be
\begin{align}
E^1 &= \sqrt{\frac{g_3(Y)}{2}}dY\qquad E^2 =\sqrt{\frac{2}{9g_3(Y)}}(d\psi + 3A_{FS}) \cr
E^{3,4,5,6} &= \sqrt{2Y}\, e^{1,2,3,4} \,.
\end{align}
The function $g_3(Y)$ is explicitly given by
\be
g_3(Y) = \frac{36Y^2\, e^{\frac{6Y}{k}}}{k(-k^2+e^{\frac{6Y}{k}}(k^2-6kY+18Y^2 )}\,,
\ee
We choose six dimensional Weyl spinors as in appendix \ref{spinors}. 
The Dirac equation for the positive chirality spinor takes the schematic form
\begin{align}
S_i(\theta)T_i(\chi) L_i(Y)\cdot G_i(Y) + G_j(Y)T_j(\chi) N_i(\theta)\cdot S_j(\theta) + G_k(Y) S_k(\theta) M_i(\chi)\cdot T_k(\chi) &= 0\,.
\end{align}
where the indices $i, j, k$ range over the four components of a chiral
spinor. Here, $L_i$, $M_i$ and $N_i$ denote differential operators
depending on the single variable in parenthesis. As in the four-dimensional background, we impose
the stronger condition that $L\cdot G = N\cdot S = M\cdot T =
0$\,. The positive chirality solution is then given, up to the
phase factor $e^{in\psi + i p \varphi+ i m\phi}$, by:
\be
\Psi_+ = \begin{pmatrix}
\frac{C_5}{\sqrt{Y}}\, e^{\frac{3Yn}{k} - wY} \left(\frac{36 Y^2}{kg_3(Y)}\right)^{-\frac{1}{4}+\frac{n}{2}} (\sin\chi)^{-2+2p} (\cos\chi)^{-\frac{1}{2}-2p-3n}
(\sin\theta)^{-\frac{1}{2}-p}(\tan\frac{\theta}{2})^{m} 
\cr
\frac{C_6}{\sqrt{Y}}\, e^{\frac{3Yn}{k} - wY} \left(\frac{36 Y^2}{kg_3(Y)}\right)^{-\frac{1}{4}+\frac{n}{2}} (\sin\chi)^{-2-2p} (\cos\chi)^{-\frac{1}{2}+2p+3n}
(\sin\theta)^{-\frac{1}{2}+p}(\tan\frac{\theta}{2})^{-m} 
\cr
\frac{C_7}{ Y}\, e^{-\frac{3Yn}{k} + wY} \left(\frac{36 Y^2}{kg_3(Y)}\right)^{-\frac{1}{4}-\frac{n}{2}} (\sin\chi)^{-1-2p} (\cos\chi)^{-\frac{1}{2}+2p+3n}
(\sin\theta)^{-\frac{1}{2}-p}(\tan\frac{\theta}{2})^{m} 
\cr
C_8e^{-\frac{3Yn}{k} + wY}\, \left(\frac{36 Y^2}{kg_3(Y)}\right)^{-\frac{1}{4}-\frac{n}{2}} 
(\sin\chi)^{-1+2p} (\cos\chi)^{-\frac{1}{2}-2p-3n}
(\sin\theta)^{-\frac{1}{2}+p}(\tan\frac{\theta}{2})^{-m}
\end{pmatrix}
\nonumber 
\ee 
We summarize the asymptotics of the radial function:
\be
Y^3 \quad\xleftarrow{Y\rightarrow 0}\frac{Y^2}{g_3(Y)} \xrightarrow{Y\rightarrow \infty}\quad \frac{kY^2}{2} \,,
\ee
and note the measure factor:
\be
\sqrt{\det g} =\frac{1}{6} Y^2 \cos \chi \sin \theta \sin^3 \chi.
\ee
{From} the flat space limit and the fermionic nature of the target
space wave-functions, we conclude that $3n$ needs to be strictly
half-integer, while $2m$ and $2p$ are integers of opposite
parity. See appendix \ref{Hopf} for details.

For negative momentum $n$, by analyzing the radial profile near the tip,
we conclude that $C_5=C_6=0$. We therefore focus
 on the $7$ and
$8$ components. For the 7 component, we find that
\be
C_7\ne 0 \Rightarrow -p \pm m \ge \frac{1}{2}\quad\text{and}\quad 3n+ 2p \ge 0 \,,
\ee
which are contradictory, implying that $C_7=0$. For the 8 component, we find
\be
C_8\ne 0 \Rightarrow p \pm m \ge \frac{1}{2}
\quad \text{and}\quad  +3n+ 2p \le 0 \,.
\ee
We can think of the quantum number $m$ as filling out a spin $p-1/2$ multiplet.
When we refine the counting through various quantization conditions,
and combine all constraints, we find that, for the $8$ component, the
degeneracy of zero modes for a given value of momentum $n$ is given by
\begin{align}
D(n) &= \sum_{p=1/2,1, \dots}^{-\frac{3n}{2}-\frac{1}{4}} 2p=  \sum_{r=1}^{-3n-\frac{1}{2}} r =\frac{(-3n-\frac{1}{2}) (-3n+\frac{1}{2})}{2}\,. 
\end{align}
When we analyze the positive momentum $n$ states, we find no solutions.
For the negative chirality spinor the wave function is (up
to the phase factor $e^{in\psi+ip\lambda + im\phi}$) of the form:
\begin{equation}
\Psi_-= 
%e^{in\psi+ip\lambda + im\phi}
\begin{pmatrix}
C_1\, e^{\frac{3Yn}{k} -wY}\left(\frac{36 Y^2}{kg_3(Y)}\right)^{-\frac{1}{4}+\frac{n}{2}}
(\sin\chi)^{-1-2p}(\cos\chi)^{-\frac{1}{2}+2p+3n}
(\sin\theta)^{-\frac{1}{2}-p}(\tan\frac{\theta}{2})^m \cr
\frac{C_2}{Y}\, e^{\frac{3Yn}{k} -wY}\left(\frac{36 Y^2}{kg_3(Y)}\right)^{-\frac{1}{4}+\frac{n}{2}}
(\sin\chi)^{-1+2p}(\cos\chi)^{-\frac{1}{2}-2p-3n}
(\sin\theta)^{-\frac{1}{2}+p}(\tan\frac{\theta}{2})^{-m} \cr
\frac{C_3}{\sqrt{Y}}\, e^{-\frac{3Yn}{k} +wY}\left(\frac{36 Y^2}{kg_3(Y)}\right)^{-\frac{1}{4}-\frac{n}{2}}(\sin\chi)^{-2+2p}(\cos\chi)^{-\frac{1}{2}-2p-3n}
(\sin\theta)^{-\frac{1}{2}-p}(\tan\frac{\theta}{2})^m\cr
\frac{C_4}{\sqrt{Y}}\, e^{-\frac{3Yn}{k} +wY}\left(\frac{36 Y^2}{kg_3(Y)}\right)^{-\frac{1}{4}-\frac{n}{2}}(\sin2\chi)^{-2-2p}(\cos\chi)^{-\frac{1}{2}+2p+3n}
(\sin\theta)^{-\frac{1}{2}+p}(\tan\frac{\theta}{2})^{-m}
\end{pmatrix}
\end{equation}
One notices that the angular constraints from the third and fourth
components lead to contradictory requirements on the quantum number
$p$, enforcing $C_3=C_4=0$. We are then led to the constraint that
the momentum $n$ be positive for negative chirality spinors and find that only the first component can be
non-zero. From the angular variables we first of all find the
constraints
\be
p+\frac{1}{2}\le m \le-p-\frac{1}{2}
\ee
with $p < -\frac{1}{2}$. Once again, the variable $m$ takes values in a spin $|p|-\frac{1}{2}$ representation, leading to a $2|p|$ degeneracy factor. The positivity of the $\cos\chi$ exponent leads to the constraint
\be
p > \frac{1}{4} -\frac{3n}{2}\,,
\ee
which gives rise to an angular degeneracy of
\begin{align} 
D(n) =\sum_{p=\frac{1}{4}-\frac{3n}{2}}^{-\frac{1}{2}} 2|p| =\sum_{r=\frac{1}{2}-3n}^{-1}|r|= \frac{(3n-\frac{1}{2})(3n+\frac{1}{2})}{2}\,.
\end{align}
{From} the behavior at infinity, we find some further constraints on 
the quantum number $n$. We get (with the same regularization as in two
and four dimensions):
\begin{align}
& kw < 3n \le -\frac{3}{2} \quad \mbox{or}\quad kw \ge 3n \ge \frac{3}{2}\,.
\end{align}
These lead to the following expression for the final index sum:
\begin{align}
Z_6 &= \left(\sum_{w<0}\sum^{3n=-3/2}_{ {[[} kw - \epsilon {]]}}-
 \sum_{w>0}\sum_{3n=3/2,5/2,\dots}^{ {[[} kw {]]}}\right)
  \frac{(3n-\frac{1}{2}) (3n+\frac{1}{2})}{2} y_1^n y_2^w\cr
&= \left(\sum_{w<0}\sum^{m=-2}_{ {[[} kw  - \epsilon {]]}-1/2}-
 \sum_{w>0}\sum_{m=1,2,\ldots}^{ {[[} kw {]]}-1/2 } \right)
  \frac{m(m+1)}{2} y_1^m y_2^w.
\end{align}
where we identified $m=3n-1/2$.

\subsection{The physics underlying the index formulas}\label{physics}

We would like to summarize and highlight some of the common features
of the results we have obtained so far. Elementary physical reasoning
will give some extra insight into the results of the explicit counting.
The general form of the answer for the
index of the Dirac operator, for a fixed winding
sector, can be written in the form,
\begin{align}\label{generalindex}
Z_{2N}(w) = \begin{cases}
\sum_{m=M(w)}^{-1}\ D(m,N) y^{n} & \qquad  \text{for}\qquad w<0\\
(-1)^N \sum_{m=1}^{M(w)} \ D(m,N)y^{n} & \qquad \text{for} \qquad w>0\,,
\end{cases}
\end{align}
where $M(w)$ is a winding dependent bound on the summation range for the
quantum number $m$.
The full index is obtained by summing over all possible values of the
winding $w$. The degeneracy factor $D(m,N)$ was obtained by studying the
angular part of the solutions of the Dirac equation in detail. We
found that
\begin{equation}
  D(m,N)=\begin{cases}
1, & \text{for}\quad N=1\\
m, & \text{for}\quad N=2\\
\frac{1}{2}m(m+1)&\text{for}\quad N=3\,.
  \end{cases}
\end{equation}
In fact, there is another route to these degeneracy factors that
simultaneously provides its generalization. When we study the
background (\ref{dmetric}), and concentrate on the compact directions
along the $\mathbb{CP}^{N-1}$, we can imagine performing a
Kaluza-Klein reduction on the circle parametrized by $\psi$, near the
tip (where there is no winding contribution to the energy). That gives
rise to a gauge field $N A_{FS}$ on the complex projective space, and
given that our fermion has momentum $n$, which turns into an electric
charge, we obtain a charged massless fermion on $\mathbb{CP}^{N-1}$
with charge-magnetic field product equal to $n N$. The Dirac zero
modes on the whole space correspond to Dirac zero modes in this
compact slice as well, such that we must determine the degeneracy of
the lowest Landau level of this generalized quantum Hall system
\cite{Karabali:2002im}. This degeneracy is fixed by an index theorem
on $\mathbb{CP}^{N-1}$ with magnetic field \cite{Dolan:2003bj}. The
index is given by the integral over the manifold of a power of the
generating (Fubini-Study) line bundle $L$ and the A-roof genus:
\begin{align}
D (nN-\frac{N}{2}+1,N) &= \int_{\mathbb{CP}^{N-1}} \mbox{ch}(L^{nN})
\hat{A} (\mathbb{CP}^{N-1})
 \cr
&= \frac{1}{(N-1)!} (nN-\frac{N}{2}+1) ( nN-\frac{N}{2}+2) \dots
 (nN-\frac{N}{2}+N-1). \cr
\end{align}
We must identify $m=nN-N/2+1$ and indeed find a perfect match between
the degeneracy of states in the lowest Landau level and the number of
Dirac zero modes.  %The identification can be motivated as
%follows. 
A corollary of the matching between
the degeneracy and the index theorem is that we have only solutions of
one chirality (at fixed winding number $w$). An alternative route to
finding this degeneracy is through a careful analysis of
representations of $U(1) \times SU(N-1) \subset SU(N)$
\cite{Karabali:2002im}. 
Near the tip of the geometry, we effectively analyze
the zero modes of a fermionic particle carrying a charge $nN$ and
charged under the $N-1$
%$U(1)$ 
Cartan generators of
the $SU(N)$ isometry subgroup of $\mathbb{CP}^{N-1}$. Labelling the $U(1)$
eigenvalues by non-negative $l_i$, (where $i\, \in \, \{ 1,2,
\dots,N-1 \} \,$) we recall that the wave function at a given energy
eigenvalue is characterized by the $U(1)$ gauge field charge $nN$ and the
total angular momentum  $l=\sum_{i=1}^{N-1} l_i$ \cite{Kirchberg:2004za}. For a given
total angular momentum $l$ there are ${l+ N-2
  \choose N-2}$ wave functions.
The angular momentum of ground states is bounded above in terms of the $U(1)$
gauge charge. The upper bound is $m-1=nN-N/2$. 
 We have thus acquired a better understanding
of the origin of the degeneracy of the Dirac zero modes, and the
physics of a radial cross section near the origin.

On the other hand, the limits on the variable $m$, denoted $M(w)$, are
determined by the radial problem, and in particular the coefficient of
the radial exponential in the wave function.
 This can be obtained for the
general case by studying the Dirac equation near radial infinity. In
this limit, one can define a radial coordinate $\rho \approx
\frac{Y}{k}$and write the metric in the form
\be
\frac{ds^2}{k} \approx d\rho^2 + \frac{R^2}{N^2} (d\psi+N A_{FS})^2  + 2\rho\ ds^2_{FS} \,.
\ee
We have kept a general radius $R$ for later purposes. For now we have $R=1$.
The gauge field is of the form
\be
A = \frac{w}{f(Y)}(d\psi + N A_{FS})\,,
\ee
where the function $f$ is given by
\begin{equation}
  f(Y)= %\begin{cases}
\frac{N}{2} g_N(Y) \,. %& \text{for type II}\\
%h_N(Y), & \text{for heterotic}\,.
  %\end{cases}
\end{equation}
Writing out the Dirac equation for a charged fermion in these
coordinates we find that it takes the following form for the
individual components of the fermion
\be
\left[\frac{\p}{\p \rho} \pm  \frac{1}{R}\left(i\frac{\p}{\p\psi} + A_{\psi} \right)\right] \Psi_a = 0 \,.
\ee
With $\psi$-momentum equal to $n$, we obtain the asymptotic solution
\be
\Psi_a = e^{in\psi}\, e^{\pm\frac{\rho}{R}(\frac{w}{f(\infty)}-n)} \,.
\ee
Therefore, we find normalizable modes whenever
\be
|n| < \left|\frac{w}{f(\infty)}\right| \,.
\ee
In this section, we have $f(\infty) = N/k$.
The bounds on 
$m=nN-N/2+1$ therefore read:
\be
|m+ \frac{N}{2}-1| < \left| kw \right| \,.
\ee
This matches with what we obtained in previous sections. Besides an
explicit description of the Dirac zero modes, we now also have an
intuitive understanding of the counting formulas.

Using this improved understanding, we can generalize our result for any value
of $N$. First of all, we observe that we have that $nN$ is a strict half-integer
for $N$ odd, while it is an integer for $N$ even. As a consequence, the number $m$
is always integer. Using this fact, and the previous results, we
find the partition sum for $N$ even:
\be
Z_{2N} = \left( \sum_{w<0}\sum^{m=-\frac{N}{2}+1}_{ {[} kw - \epsilon {]} + 1 - \frac{N}{2}}
- \sum_{w>0}\sum_{m=1,2,\dots}^{ {[} kw {]} + 1 - \frac{N}{2}} \right)
D(nN-\frac{N}{2}+1,N) \,,
\ee
while for $N$ odd:
\be
Z_{2N} = \left(\sum_{w<0} \sum^{m=-\frac{N}{2}+\frac{1}{2}}_{ {[[} kw - \epsilon {]]} + 1 - \frac{N}{2}}
- \sum_{w>0}\sum_{m=1,2,\dots}^{ {[[} kw {]]} + 1 - \frac{N}{2}} \right)
D(nN-\frac{N}{2}+1,N) \,.
\ee
For $N$ odd, we obtain contributions of different chiralities,
depending on the sign of the momentum, while for $N$ even, we obtain
contributions only of a given chirality, but we need to take care of
the fact that the degeneracy factor flips sign when going from
positive to negative values of the momentum. 

We note that the degeneracy factor is a polynomial in the quantum number $m$ of degree $N-1$.
If we introduce a fugacity corresponding to the quantum number $m$, and suppose
for instance that its norm is smaller than one, then the double sum over the winding and momentum will
give rise to a pole in the fugacity of order $N$. We will explicitly perform such a resummation
for $N=1$ in section \ref{con}.

\section{Duality rotated asymptotically linear dilaton spaces}
\label{het}

In this section, we revisit the Dirac index calculation in a
supergravity background which is either a background with asymptotic
linear dilaton or a Euclidean Taub-NUT geometry. We will use a
different technique to add a potential to the moduli space. Our logic
is to start with a background which is a solution to the
low-energy effective supergravity arising from (for instance) a
heterotic string theory. The backgrounds only have a non-trivial
metric and possibly a linear dilaton.  We then perform a duality
rotation on the isometric direction and an internal direction
associated to a gauge field. We  thus generate new backgrounds of
supergravity that are of independent interest. The main point though is that
the rotation introduces a gauge field in the background, which renders
the counting problem for charged bound states well-defined.

\subsection{Duality rotated backgrounds}
In the first part of this section, we recall how to generate the new
supergravity backgrounds through duality rotations.  We consider the
asymptotically linear dilaton background in \eqref{dmetric} as being
part of a heterotic string background. As explained in
\cite{Johnson:1994ek}, based upon the results of \cite{S1, HS2}, one
can supplement such a $2N$-dimensional space-time with an extra chiral
coordinate $X$, which allows one to introduce a background abelian
gauge field.  We will generate a new solution by doing an $O(2)$
transformation, although more general duality transformations are possible.
%C

The duality rotation is most easily performed in the following way: encode the metric $G$, the gauge field $A$ and the anti-symmetric tensor $B$ into a $4N+1 \times 4N+1$ dimensional matrix
\be
{\cal M} = \begin{pmatrix}
K_-^TG^{-1}K_-& K_-^T G^{-1} K_+ & - K_- G^{-1}A \cr
K_+^TG^{-1}K_-& K_+^T G^{-1} K_+ & - K_+ G^{-1}A\cr
-A^TG^{-1}K_- & -A^TG^{-1}K_+ & A^TG^-{1}A 
\end{pmatrix}\,.
\ee
Here $T$ denotes transposition and the auxiliary $2N\times 2N$ dimensional matrices $K_{\pm}$
are defined by
\be
(K_{\pm})_{\mu\nu} =-B_{\mu\nu}-G_{\mu\nu}-\frac{1}{4}A_{\mu}A_{\nu}\pm \eta_{\mu\nu}\,.
\ee
Since the backgrounds we work with are Euclidean we will work with the Euclidean signature matrix $\eta=\mathbb{I}_{2N}$. New solutions to low-energy heterotic supergravity actions are  generated by performing a duality rotation
\be
{\cal M} \rightarrow {\cal M}' = \Omega {\cal M}\Omega^{T}	
\ee
where $\Omega$ is a rotation matrix.
We choose a duality rotation in an $SO(2)$ subgroup:
\be
\Omega = \begin{pmatrix}
\mathbb{I}_{2N-1} & 0 & 0 \cr
0 & \gamma & \sqrt{1-\gamma^2}\cr
 0 &-\sqrt{1-\gamma^2} &\gamma
\end{pmatrix},
\ee
where $\gamma$ is a rotation parameter that takes values in
$[0,1]$. The entries that are non-zero are chosen to be along one of
the isometry directions of the manifold $M_d$
and an internal
direction associated to the gauge field $A$.

The new metric, anti-symmetric tensor and gauge field can be unpackaged
from the entries of the matrix ${\cal M}'$. Finally, the
dilaton in the original and new solution are related through the duality invariant
combination
\be
e^{-2\Phi'} \det{G'}= e^{-2\Phi} \det{G} \,.
\ee
In the following, we apply these duality transformations to the
asymptotically linear dilaton spaces in various dimensions.
\subsubsection{Asymptotically linear dilaton dyons}

Let us consider the effect of the duality rotation on the asymptotically linear dilaton backgrounds given in
equation (\ref{dmetric}). We perform the $O(2)$ duality rotation involving the $\psi$-direction and an internal direction. The resultant metric and dilaton takes the following form 
\begin{align}\label{hetmetanddil}
ds^2 &= \frac{g_N(Y)}{2}dY^2 + \frac{N^2\, \b^2 g_N(Y)}{2\, h_N(Y)^2}\left(\frac{d\psi}{\b} + N \, A_{FS}\right)^2 + 2Y ds^2_{FS}  \cr
e^{-2\Phi} &= e^{\frac{2NY}{k}}\, \left(\frac{2h_N(Y)}{N^2g_N(Y)}\right)^2 \,,
\end{align}
where $\b \in [0,1]$. 
%C
We defined the function
\be
h_N(Y) = 1-\b + \frac{N^2\b}{2} \, g_N(Y)\,.
\ee
The other closed string fields that result from the duality transformation are the
gauge field $A$ and the NS-NS two-form $B$:
\begin{align}\label{hetAB}
A &= -2\sqrt{\frac{1-\b}{\b}}d\psi +\frac{2\sqrt{(1-\b)\b}}{h_N(Y)}\left(\frac{d\psi}{\b} +  NA_{FS}\right) \cr
B&= \frac{N\b (1-\b)}{h_N(Y)}\frac{d\psi}{\b} \wedge A_{FS}\,.
\end{align}
From the form of the rotated metric, gauge field and B-field, one can
see that, if we rescale the coordinate $\psi$ by $\beta$, the new
coordinate allows one to write the fibration over $\mathbb{CP}^{N-1}$
in a manner identical to the linear dilaton fibrations discussed in appendix
B. In other words, the metric and gauge field of these duality rotated
backgrounds differ from the background in \eqref{dmetric} only in the
detailed form of the radial functions that appear in the
fibration. This will have important consequences %as we see 
%C
below. We
will once again denote the new coordinate by $\psi$ and it has
periodicity $2N\pi$. This choice of periodicity renders the duality rotated backgrounds
genuinely new. We moreover drop the first, Wilson line term in the gauge field in equation
(\ref{hetAB})
%C
 -- the background
remains a supergravity solution.
%C

\subsection{Charged Dirac Indices}

In the following we analyze the Dirac index of a fermion in the
duality rotated metric that we generated, coupling with strength $e$
to the gauge field $A$ generated by the duality rotation.\footnote{We
  will not limit ourselves to embeddings of this model in string
  theory. In that context, the charge $e$ would be
  quantized. The supersymmetric quantum mechanics model corresponding
  to a manifold with gauge bundle \cite{Friedan:1983xr} can in any
  case be thought off as the dimensional reduction of a heterotic
  non-linear sigma-model.}  The Dirac equation we solve includes a 
coupling of the charged fermion to the gauge field:
\be
\gamma^{\mu}(\p_{\mu}+\frac{1}{4}\omega_{\mu}^{ab}\gamma_{ab} -i e A_{\mu})\Psi=0\,.
\ee
We discuss the asymptotic linear dilaton models for general $N$.
We follow the intuitive 
discussion in subsection \ref{physics}. The
physics near the origin $Y=0$ is the same as before : we project onto
the lowest Landau level
 on the projective space $\mathbb{CP}^{N-1}$. The
asymptotics of the radial problem at $Y\rightarrow \infty$ is slightly
different\footnote{We 
rescaled the electric charge $e$ to $\frac{e}{2\sqrt{\b (1-\b)}}$.}:
\begin{align}
\frac{ds^2}{k} &\approx \frac{dY^2}{k^2} + \frac{N^4\b^2}{k^2h_N(\infty)^2}(d\psi +N A_{FS})^2 + \frac{2Y}{k}ds_{FS}^2\,, \cr
eA&\approx\frac{e
%2\sqrt{\b(1-\b)}
}{h_N(\infty)}(d\psi + NA_{FS})\,.
\end{align}
If we make the identifications
\begin{align}
w & =  e
%\sqrt{\b(1-\b)}  
\cr
 f(Y) &= h_N(Y) \cr
 R &= \frac{N^2\b}{k\, h_N(\infty)}
\quad
\text{with}\quad h_N(\infty) = 1-\b +\frac{\b N^2}{k}
\end{align}
we see that the asymptotic analysis of the restrictions on the momentum of the bound state
runs exactly as in subsection \ref{physics}. 
One can thus write down the index for these backgrounds
for a given value of the electric charge:
\begin{align}\label{indexgeneral}
Z_{2N}(e)  = \begin{cases}
\qquad \quad \sum_{n=\frac{1}{2}}^{ {[} \frac{e
%2\sqrt{\b(1-\b)}
}{h_N(\infty)} -\epsilon{]}} \, D(nN-\frac{N}{2}+1,N) y^n & \qquad\text{for}\qquad e >0\\
(-1)^N\sum_{n={[}\frac{e
%2\sqrt{\b(1-\b)}
}{h_N(\infty)} {]}}^{-\frac{1}{2}}\, D(nN-\frac{N}{2}+1,N) y^n & \qquad\text{for}\qquad e <0 \,.
\end{cases}
\end{align}
Here we have used the degeneracy function $D$ that counts the index of
the Dirac operator on $\mathbb{CP}^{N-1}$ with magnetic field. For odd $N$ we have to replace
the square brackets appropriately.

There are some subtle differences between this class of models and the
ones we studied earlier. In particular, for the higher dimensional
heterotic backgrounds it turns out that the Dirac equation for the components are not
explicitly integrable in terms of known special functions. 
In order to illustrate these differences, we show how to proceed with
the explicit calculation of the zero modes in the four-dimensional
example. We confirm that the counting of zero modes leads to
equation \eqref{indexgeneral} for the case 
$N=2$. A similar analysis governs the case of 
dimension $2N$.

\subsubsection{Four-dimensional asymptotically linear dilaton dyon}
For the calculation of the index on the four-dimensional heterotic dyon, we choose the following vierbein: 
\begin{align}
E^1 &= \sqrt{\frac{g_2(Y)}{2}}dY \qquad E^2 = \frac{\sqrt{2g_2(Y)}\b}{h_2(Y)}( d\psi + \cos\theta d\phi)\cr
E^3 &= \sqrt{\frac{Y}{2}} d\theta \qquad E^4 = \sqrt{\frac{Y}{2}}\sin\theta d\phi \,.
\end{align}
The gauge field is given by
\be
eA = \frac{e
%2\sqrt{\b(1-\b)}
}{h_2(Y)}(d\psi + \cos\theta d\phi) \,.
\ee
The Dirac equation is solved as before and the positive
chirality wave functions are
\be
\Psi_+ = e^{i n \psi+im\phi}\begin{pmatrix}
G_3(Y)(\sin\theta)^{-\frac{1}{2}+ n}(\tan\frac{\theta}{2})^{-m}\cr
G_4(Y)(\sin\theta)^{-\frac{1}{2}- n}(\tan\frac{\theta}{2})^{m}
\end{pmatrix}\,.
\ee
The radial wave functions
satisfy the equations
\begin{align}
\frac{dG_a(Y)}{dY} + W(Y) G_a(Y) &= 0\quad\text{with}\quad a=3,4 \,.
\end{align}
where
\begin{align}
W_3(e, n, Y)& = \frac{e
%2\sqrt{\b(1-\b)}
}{2\b} - \frac{n}{2 \beta} h_2(Y) + \frac{1}{4Y}+ \frac{g_2'(Y)}{4g_2(Y)}-\frac{k\beta g_2'(Y)}{h_2(Y)} + \frac{(1-\beta)}{4Yh_2(Y)} \cr
\text{and}\quad W_4(e, n, Y)&=W_3(-e, -n, Y)\,.
%& =e\sqrt{\frac{1-\b}{\b}} + \frac{n}{2k} h_2(Y) +\frac{1}{4Y}+ \frac{g_2'(Y)}{4g_2(Y)}-\frac{k\beta g_2'(Y)}{h_2(Y)} + \frac{(1-\beta)}{4Yh_2(Y)} \,.
\end{align}
All but the last term are explicitly integrable in terms of known functions. The solution is of the form 
\begin{align}
G_3(Y) &= e^{\int dY \frac{(1-\b)}{4Yh_2(Y)}}\, 
%e^{\frac{2 n Y}{k}}
\sqrt{\frac{h_2(Y)}{Y}}\, 
e^{-\frac{Y}{2 \b}\left(e
%2\sqrt{\b(1-\b)}
-n (1-\b+\frac{4\b}{k})\right)}\,
\left(\frac{8Y}{kg_2(Y)}\right)^{\frac{1}{4}+\frac{ n}{2}}\cr 
G_4(Y) &= e^{\int dY \frac{(1-\b)}{4Yh_2(Y)}}\,
% e^{-\frac{2 n Y}{k}}
\sqrt{\frac{h_2(Y)}{Y}}\, 
e^{\frac{Y}{2 \b}\left(e
%2\sqrt{\b(1-\b)}
-n (1-\b+\frac{4\b}{k})\right)}\,
\left(\frac{8Y}{kg_2(Y)}\right)^{\frac{1}{4}-\frac{ n}{2}}
\,.
\end{align}
%
%$G_3$ is obtained by flipping the sign of $n$ and $e$. 
By analyzing the profile of the function $h_2(Y)$, 
one can check that the non-integrable term leads to a finite
correction to the wave-function, well-behaved near the tip at $Y=0$
and near radial infinity. It is immaterial for the purposes of
counting the bound states. There are no normalizable modes with
negative chirality. The index partition sum is calculated precisely as
before with the result:
\begin{align}
Z_4^{het}(e) &= \left(\sum_{2n=1}^{ {[} 
\frac{2 e
%2\sqrt{\b(1-\b)}
}{h_2(\infty)} -\epsilon{]}}
  -\sum^{-1}_{2n={[}\frac{2 e
%2\sqrt{\b(1-\b)}
}{h_2(\infty)}{]}} \right)\, 2n\, y^{n}\,.
\end{align}
where we have identified $h_2(\infty)=1-\b+\frac{4\b}{k}$. This indeed coincides with the formula in equation \eqref{indexgeneral} which was derived based on general considerations.

\section{Conclusions and Future Directions}
\label{con}
In this paper, we computed the indices of supersymmetric quantum
mechanical models associated to asymptotically linear dilaton
spaces. We used two ways of regularizing a marginal bound state
problem. Firstly, through the introduction of a potential consistent
with supersymmetry and associated to a Killing isometry in target
space (either through winding, or through rotating an orthogonally wound
string). Secondly,
through a supergravity duality transformation that introduces a gauge
field, and coupling the space-time spinor to the gauge field. We
calculated the indices by explicitly solving the massless
Dirac equation. We argued that we can compute the indices by
solving linear equations, which leads to a technical improvement on
the counting of bound states for instance on Euclidean Taub-NUT with
self-dual gauge field. The physical intuition behind the counting
formulas is that the zero modes have a degeneracy equal to that of the
lowest Landau level on $\mathbb{CP}^{N-1}$.  It would be interesting
to employ generalized index theorems on non-compact asymptotically
linear dilaton spaces to  check our results using a
topological index calculation.  We believe the supersymmetric quantum
mechanical models that we defined and partially studied in this paper
are interesting by themselves, and sufficiently versatile to find numerous
applications in higher dimensional supersymmetric theories such as
supersymmetric field theory or string theory.  In the following, we
wrap up by pointing out potential applications of our results in two
specific contexts.

\subsection{Mock Modular Forms}

We would like to illustrate that the quantum mechanics index
calculation is sufficient to reproduce the mock modular form that
features in the elliptic genus of the $N=2$ superconformal cigar
conformal field theory in $1+1$ dimensions 
\cite{Eguchi:2004yi, Troost:2010ud, Eguchi:2010cb, Ashok:2011cy}.
 It is
important to first 
understand what aspects of the conformal field theory are captured by the
super quantum mechanics obtained via the Scherk-Schwarz reduction. It
is well known that the $N=2$ cigar coset conformal field theory can be written as a
product of free fermions and bosons. It is only the zero modes
of the bosonic fields that are captured by the super quantum
mechanics.

We have determined the number of space-time positive and negative chirality
zero-modes of a space-time fermion. The space-time fermion can be
thought of as arising from the quantization of right-moving fermionic
zero-modes of the worldsheet conformal field theory, and bosonic zero modes.
To obtain a space-time fermion, we would take the left-movers 
in the NS sector. In what follows, we will prefer to think about
Ramond-Ramond sector states in the conformal field theory. These
correspond to space-time bosons, and we will accordingly shift the
momentum $n$ from half-integer to integer.

The elliptic genus of the coset conformal field theory is defined as
follows (see e.g. \cite{Kawai:1993jk}):
\be
\chi_{cos} = \text{Tr}(-1)^{F}q^{L_0-\frac{c}{24}}\bar{q}^{\bar{L}_0-\frac{c}{24}}z^{J_0} y^P \,,
\ee
where $L_0$ denotes left-moving conformal dimension, $J_0$ left-moving R-charge, and
$P$ a global $U(1)$ charge (e.g. momentum).
In order to lift the index calculation and complete it to an elliptic
genus what we would like to do is to express the R-charge and
conformal dimensions for both left and right moving states of the
Ramond-Ramond sector in terms of the momentum and winding around the
asymptotic circle $\psi$. The idea then is to assign appropriate
weights (dependent on the modular parameter $q$ and the R-charge chemical potential
$z$) to the variables $y_1$ and $y_2$ in the index result
\eqref{cigarZ2} that would allow
 us to read off the contribution of
the quantum mechanics to the elliptic genus of the conformal field
theory.

The conformal algebra of the cigar conformal field theory has been
discussed in detail in for instance \cite{Israel:2004jt}.  Since the
quantum mechanics captures the contribution of the bosonic zero modes,
their contribution to the conformal dimension and R-charge can be read
off to be
\begin{align}
L_0^{QM} - \frac{1}{4k}  &= \frac{(n-kw)^2}{4k} + \frac{p_{rad}^2}{2}\cr
J_0^{QM} &= \frac{n-kw}{k} \,.
\end{align}
Here $p_{rad}$ is the radial momentum along the $Y$ direction. The right moving conformal dimension is similarly given by
\be
\bar{L}_0^{QM} - \frac{1}{4k} = \frac{(n+kw)^2}{4k} + \frac{p_{rad}^2}{2}\,.
\ee
We now impose the right-movers to be in the ground state in order to
calculate the elliptic genus; this automatically fixes the total right-moving
conformal dimension $\bar{L}_0 =
\frac{c}{24}$. For the left movers, we have to calculate
the left-moving conformal dimension $L_0-\frac{c}{24}$, which we can rewrite as the difference
between left-moving and right-moving conformal dimension $L_0-\bar{L}_0$. Using the
explicit expressions for these charges, we find that
\be
L_0 - \bar{L}_0 = -nw \,.
\ee
Using this and the expression for the R-current in terms of the
momentum and winding quantum numbers, we can now assign appropriate
weights in order to calculate that part of the elliptic genus of the
cigar conformal field theory that arises from the super quantum
mechanics.\footnote{The sign of the momentum is conventional. We have flipped it here for easier
comparison with  \cite{Troost:2010ud, Ashok:2011cy}. We restrict in this section to
integer levels $k$.} We obtain
\be\label{Z2}
Z_2 
=  \left(\sum_{n \le -1, n+kw \ge 0} - \sum_{n \ge 0, n+kw \le -1 }\right) z^{ \frac{kw-n}{k}} q^{-nw} y^n \,.
\ee
We shifted the momentum $n$ downward by $1/2$ to take into
account the fact that we spectrally flowed from the NS sector to the R
sector for the left-movers.

We shall now show that this results captures the non-trivial factor of
the holomorphic part of the full elliptic genus.  To show this, we
start with the holomorphic contribution \cite{Eguchi:2004yi} to the
cigar coset elliptic genus from
\cite{Troost:2010ud, Eguchi:2010cb, Ashok:2011cy}:
\be
\chi_{cos,hol}
= \frac{1}{k} \sum_{\gamma,\delta \in \mathbb{Z}_k}
e^{\frac{2 \pi i \gamma \delta}{k}}
   \frac{i \theta_{11}(\tau, \a)}{\eta^3}
    \sum_{m \in \mathbb{Z}} \frac{
q^{\frac{(km+\gamma)^2}{k}} z^{2 \frac{km+ \gamma}{k}}    }{1-z^{\frac{1}{k}}
q^{m+ \frac{\gamma}{k}} e^{\frac{2 \pi i \delta}{k}} } y^{-(\gamma+km)},\label{doublesum}
\ee
We then follow \cite{Troost:2010ud, Ashok:2011cy} and expand the result in a particular
regime of parameters $(y,z,q)$ as a double sum:
\be
\chi_{cos,hol} 
 = 
   \frac{i \theta_{11}(\tau, \alpha )}{\eta^3} 
\left( \sum_{n \le 0, n+kw \ge 0}- \sum_{n  \ge 1, n+kw \le -1}
\right)
z^{\frac{kw-n}{k}}  q^{-nw} y^n. \label{coshol}
\ee
We see several differences between formulas (\ref{Z2}) and
(\ref{coshol}). As mentioned earlier, in the quantum
mechanics, 
we have not taken into account the left-moving oscillator contributions of fermions
or bosons, nor the degeneracy of the left-moving worldsheet vacuum.
Therefore, there is a factor
$i \theta_{11}(\tau,\alpha)/\eta^3$ that needs to be supplemented to equation
(\ref{Z2}).
Secondly, we see that we have taken a different scheme 
to divide $n=0$ zero modes into two sets. The scheme is a matter
of choice. The expressions (\ref{Z2}) and (\ref{coshol}) differ
by a theta-function that transforms well under modular and elliptic transformations.

Note also that if we now want to address the full elliptic genus
problem including the non-holomorphic contributions, we need to study one more aspect of the
right-moving super quantum mechanics. Indeed, the non-holomorphic
contribution contains a measure factor arising from the
difference in the density of states of primary right-moving bosons and fermions.
To determine this difference, it is sufficient to know the asymptotic form of the 
supercharge of the supersymmetric quantum mechanics. It takes the form:
\be
\tilde{Q}  \approx 
(i p_{rad} + n+ kw) \tilde{\psi} 
\, ,
\ee
where $p_{rad}$ is the radial momentum.
We can then indeed check that the
measure factor, determined in \cite{Troost:2010ud} and
 discussed in detail in \cite{Ashok:2011cy} is given by the inverse
of the asymptotic expression of the supercharge (stripped off the worldsheet fermion
operator $\tilde{\psi}$). The underlying reason for this is that the asymptotic supercharge relates
asymptotic bosonic and fermionic wave-functions, and therefore also dictates the ratio
between bosonic and fermionic reflection amplitudes. The latter in turn codes the
difference in densities of states  \cite{Akhoury:1984pt}.

Therefore, we see that in the two-dimensional cigar background, the class of
supersymmetric quantum mechanical models that we considered captures
all of the non-trivial data necessary to reconstitute the full elliptic genus of the
two-dimensional superconformal field theory, namely both the Appell-Lerch sum, and its modular
completion.

It will be interesting to apply the same idea to the higher
dimensional target spaces at our disposal and to generate new examples
of non-compact elliptic genera containing mock modular forms. The
literature on physical models for mock modular forms is growing (see
e.g.
\cite{Eguchi:2008gc, Cheng:2012tq, Alexandrov:2012au, Dabholkar:2012nd, Carlevaro:2012rz}
for recent applications). One would also like to generate models for
more exotic or new mock modular forms. The higher dimensional examples
we have are less easily lifted to full conformal field theory results,
amongst other reasons because asymptotically the $\psi$ circle remains
fibered.  Their degeneracy factors do give rise to poles of order $N$
when we re-sum but a further twist of the model may be needed to render
the degeneracies chiral on the worldsheet.  One goal could be to obtain
a conformal field theory model for the interesting double pole
Appell-Lerch sums with mock modular behavior that have been identified
in \cite{Dabholkar:2012nd} in the context of D-brane bound state
counting with application to the entropy of supersymmetric black
holes.

\subsection{Counting Domain Wall Bound States}

Euclidean Taub-NUT arises as the moduli space of monopoles in
supersymmetric Yang-Mills theory \cite{Gauntlett:1996cw, Lee:1996if} and the low energy dynamics is determined by geodesic
motion on the moduli space. A potential or self-dual gauge field
arises when we have at least a rank two gauge group, and misaligned
vacuum expectation values for adjoint scalars 
 \cite{Bak:1999da}. 
This leads to the application of index counting on
Euclidean Taub-NUT space \cite{Pope} to the counting of monopole bound
states\cite{Gauntlett:1999vc} as well as to the counting of D-brane
bound states\cite{David:2006yn}.

Interestingly it was shown in \cite{Tong:2003ik} that the cigar
background arises as the moduli space of domain walls in a three
dimensional $U(1)$ gauge theory with 8 supercharges. The simplest
setting is when the theory admits three isolated vacua, which happens
when we have three charged hypermultiplets with distinct masses (and
judicious choices for the signs of Fayet-Iliopoulos parameters). The
domain walls between the three distinct vacua preserve ${\cal
  N}=(2,2)$ supersymmetry and their low energy dynamics is described
by motion on a moduli space with cigar shaped target (when we ignore
the center of mass mode). In the infrared, the dynamics is described
by the cigar conformal field theory (or its mirror, if we approach the
dynamics through a calculation of the Liouville interaction potential)
\cite{Tong:2003ik}. This system can also be realized in string theory
in terms of D2-branes inside, and interpolating between, parallel
D6-branes in the presence of a background NSNS two-form.

It is an interesting question to explore the realization of higher
dimensional asymptotically linear dilaton spaces as moduli spaces of
(e.g. more numerous) domain walls.  Our index calculations would then
count domain wall bound states.

\section*{Acknowledgments}

We would like to thank Ghanshyam Date, Justin David, Ashoke Sen,
Nemani Suryanarayana and especially Atish Dabholkar and Sameer Murthy
for useful discussions. Our research is partly funded by the grant ANR-09-BLAN-0157-02.

\begin{appendix}

\section{First order differential equations}\label{lineareqns}

In this section we give an argument for the fact that in the models we consider,
it is sufficient to solve linear differential
equations when looking for zero modes of the twisted Dirac
operator. We start out with a  four-dimensional metric
\be\label{general4dmetric}
ds^2 = \frac{g(r)}{2} dr^2 + \frac{1}{2g(r)} (d \psi + \cos \theta d \phi)^2 +
\frac{f(r)}{2} ( d \theta^2 + \sin^2 \theta d \phi^2)
\ee
and gauge field 
\be
A=\frac{1}{2g(r)}(d\psi+\cos\theta d\phi) \,.
\ee
We choose a vierbein of the form
\be
e^1 = \sqrt{\frac{g(r)}{2}}dr\,, e^2 = \frac{1}{\sqrt{2 g(r)}}(d\psi + \cos\theta d\phi)\,, e^3 = \sqrt{\frac{f(r)}{2}} d\theta\,, e^4 = \sqrt{\frac{f(r)}{2}} \sin\theta d\phi \,. \nonumber
\ee
The resulting radial and angular differential operators entering the massless Weyl fermion equation
\begin{align}
S_3(\theta)(L_3(r) \cdot G_3(r)) + G_4(r)(N_3(\theta)\cdot S_4(\theta)) &= 0 \cr
S_4(\theta)(L_4(r) \cdot G_4(r)) + G_3(r)(N_4(\theta)\cdot S_3(\theta)) &= 0\,,
\end{align}
are:
\begin{align}
-L_3 &= \sqrt{\frac{f}{g}}\left(\frac{d}{dr}-n g+ \frac{e}{2}  \right)+ \left(\frac{g(2f'-1)-f g'}{4\sqrt{f g^3}} \right)\cr
L_4 &= \sqrt{\frac{f}{g}}\left(\frac{d}{dr}+n g- \frac{e}{2} \right)+ \left(\frac{g(2f'-1)-fg'}{4\sqrt{f g^3}} \right)\cr
N_3&=\frac{d}{d\theta}-m\csc\theta+\left(\frac{1}{2}+n\right)\cot\theta\cr
N_4&=\frac{d}{d\theta}+m\csc\theta+\left(\frac{1}{2}-n\right)\cot\theta\,.
\label{diffop}
\end{align}
These equations are equivalent to second order equations for the radial and angular functions which read as follows:
\begin{align}
N_4 N_3 S_4 &= \lambda S_4
\cr
L_3 L_4 G_4 &= \lambda G_4 \, ,
\end{align}
where $\lambda$ is a generic eigenvalue.  The operators $N_{3,4}$ are
anti-hermitian conjugate with respect to the measure $\sin
\theta$. The operators $L_{3,4}$ are hermitian conjugate with respect
to the measure $dr f^{1/2} \exp ( - \frac{1}{2} \int^r f^{-1})$.  For
example, the measure in the Euclidean Taub-NUT example is $dx (x+1)$
and in the four-dimensional asymptotically linear dilaton solution it
is simply $dY$. Assuming these measures on our space of
wave-functions, we find that the angular problem implies that $\lambda
\le 0$ while the radial problem requires $\lambda \ge 0$, thus proving
that necessarily $\lambda=0$. In turn, this implies that $N_3 S_4 = 0$
and $L_4 G_4=0$. Thus, we can restrict to solving the linear
differential equations. 

Whether the measure we chose is appropriate is a more subtle matter. Near radial infinity, the
wave functions behave exponentially, and any polynomial measure will
lead to the same conclusion about normalizability. Near the origin though,
the polynomial measure matters, but only for the smallest quantum
numbers. For those, we assume that the pattern we find for the indices
at large quantum numbers persists. In cases where one can solve
explicitly the second order differential equation, one can check the
validity of our approach even for small quantum numbers. For example
for Euclidean Taub-NUT, we can prove explicitly the validity of our
approach by comparing our results with those of Pope \cite{Pope}. This is reviewed
in appendix \ref{ETN}.\

More intuitively, the reduction to $\lambda=0$ is equivalent to a
projection onto the lowest Landau level of a charged fermion on
$\mathbb{CP}^{N-1}$ with magnetic field.
\subsubsection*{Remark on asymptotics}

For future purposes, we note an interesting aspect of the first order
differential operators (\ref{diffop}). The angular operators $N_{3,4}$ turn
out to be independent of the $\mathbb{CP}^{1}$ warp factor $f(r)$,
while the only dependence of the radial operators $L_{3,4}$ on the
metric factor $f(r)$ is encoded in the term $\frac{2 f'-1}{4 f}$. To
analyze the asymptotic behavior of the fermionic wave
function and ascertain the constraints on the
quantum numbers in the counting formula from the asymptotics
 (as in subsection \ref{physics}), we can solve the first
order system after taking the limit on the
functions $g(r)$ and $f(r)$ in the differential operators.
Consequently, two backgrounds with the same asymptotic behavior of
$g(r)$ and with the asymptotic value of $\frac{2 f'-1}{4 f}$ differing
by a subleading power of the radial coordinate, will support the same
fermionic zero modes. The index will be identical though the
backgrounds differ. This happens in the case for the counting in
Euclidean Taub-NUT done in appendix \ref{ETN} and the four-dimensional asymptotically
linear dilaton background analyzed in the main text.

\section{Fibrations over $\mathbb{CP}^{N-1}$}
\label{Hopf}

\subsection{The sphere as a circle fibration}
Let us consider the metric on an odd dimensional sphere $S^{2N-1}$. We use a simple embedding of the sphere in $\mathbb{C}^N$, parametrized by the $N$ complex coordinates $z_a$ where
\be
z_a = \mu_a\, e^{i\xi_a} \,.
\ee
The $\mu_a$ satisfy the constraint $\sum_{a}\mu_a^2 = 1$. The metric on $S^{2N-1}$ is given by
\be
ds^2_{S^{2N-1}} = \sum_{a=1}^{N} d\mu_a^2 + \sum_{a=1}^N \mu_a^2 \, d\xi_a^2 \,.
\ee
In the patch where $\mu_1\ne 0$, inhomogeneous coordinates on $\mathbb{CP}^{N-1}$ are given by
\be
Z_i = \frac{\mu_i}{\mu_1}e^{i\xi_{i1}}\qquad \text{for}\qquad i\in \{2, \ldots, N\} \,,
\ee
where $\xi_{i1}=\xi_i-\xi_1$.
The Fubini-Study metric and the associated connection one-form is given by
\begin{align}
ds^2_{FS} = \left(1+|Z_2|^2+\ldots +|Z_N|^2 \right)^{-1} \sum_{i,j=2}^{N} \left(\delta_{ij} - \frac{Z_i \bar{Z}_j}{1+|Z_2|^2+\ldots +|Z_N|^2 } \right) d\bar{Z}_i dZ_j
\end{align}
Substituting the expression for the coordinates
 $Z_i$ into the Fubini-Study metric, we find that the metric on the projective space is given by
\be
ds^2_{FS}= \sum_{a=1}^{N} d\mu_a^2 + \sum_{i=2}^{N} \mu_i^2 (d\xi_{i1})^2 - \bigg(\sum_{i=2}^N \mu_i^2\, d\xi_{i1}\bigg)^2
\ee
Subtracting the two expressions, and using the following identity:
\be
\sum_a \mu_a^2d\xi_a^2 -(\sum_{a}\mu_a^2d\xi_a)^2 -\left(\sum_{i=2}^{N}\mu_i^2(d\xi_{i1})^2 -  (\sum_{i=2}^N \mu_i^2 d\xi_{i1})^2 \right)=0\,,
\ee
we find the simple relation between the sphere metric and the metric on projective space:
\be\label{hopf1}
ds^{2}_{S^{2N-1}} = ds^2_{FS} + \bigg(\sum_{a}\mu_a^2 d\xi_a \bigg)^2 \,.
\ee
We can interpret the second term on the right as a Hopf fibration over $\mathbb{CP}^{N-1}$ as follows. The connection one-form on projective space, whose differential is the K\"ahler form, is given by
\be
A_{FS} = -\frac{i}{2} \left(1+|Z_2|^2+\ldots +|Z_N|^2 \right)^{-1}\sum_{i=2}^{N}(Z_i d\bar{Z}_i - \bar{Z}_i dZ_i) \, .
\ee
In terms of the $(\mu_i, \xi_{i1})$ coordinates, we find that
\begin{align}
A_{FS} = -\sum_{i=2}^{N}\mu_i^2\, d\xi_{i1} = d\xi_1 - \sum_{a=1}^N \mu_a^2\, d\xi_a\,,
\end{align}
In other words, the one-form, whose square equals the difference between the sphere metric and the  projective space metric is given by
\be
 \sum_{a=1}^{N}\mu_a^2\, d\xi_a = d\xi_1-A_{FS} 
\ee
One can also write this in a more symmetric form as follows:
\begin{align}
\sum_{a=1}^{N}\mu_a^2\, d\xi_a&= \frac{d\xi_1+d\xi_2+ \ldots d\xi_N}{N} + \sum_{i=2}^N \left(\mu_i^2-\frac{1}{N}\right) d\xi_{i1} \cr
&=-\frac{d\tilde{\psi}}{N} - \tilde{A}_{FS} \,.
\end{align}
We can use either $\xi_1$ or $\tilde{\psi}$ to parametrize the Hopf fibre direction. What we have therefore shown is the useful link between the sphere and the projective space metrics: 
\be
ds^2_{S^{2N-1}} = (-d\xi_1 + A_{FS})^2 + ds^2_{FS} \,.
\ee

\subsection{Asymptotically linear dilaton fibrations}

The metric of the general asymptotically linear dilaton theory is given by
\begin{align}\label{generalALD}
ds^2 &= \frac{g_N(Y)}{2}dY^2 + \frac{2}{ g_N(Y)}\left(\frac{d\psi}{N} + A_{FS}\right)^2 + 2Y ds^2_{FS}\,,  
\end{align}
The important observation that links our discussion of the circle
fibration to these metrics is the behavior of the metric near the origin
$Y\rightarrow 0$. In this limit we have
\be
ds^2 \approx \frac{dY^2}{2Y} + 2Y\left[\left(\frac{d\psi}{N}+A_{FS}\right)^2 + ds^2_{FS} \right]
\ee
Comparing, we find that we see that the metric near the tip smoothly reduces to
the flat space metric if we identify the angles
$\xi_1 = -\frac{1}{N}\psi$. This fixes the periodicity of the angle $\psi$ to be
$2\pi\, N$ since the angle $\xi_1$ in a complex two-plane has periodicity $2\pi$.

\subsubsection{Asymptotic linear dilaton fibration over $\mathbb{CP}^{1}$}
In the four-dimensional setting, we make the following choice of
coordinates on the three-sphere
\be
z_1 = \mu_1\, e^{i\xi_1}\qquad z_2= \mu_2\, e^{i\xi_2}\,,
\ee
where 
\be
\mu_1 = \cos\frac{\theta}{2}\qquad \mu_2 = \sin\frac{\theta}{2}\,.
\ee
The corresponding inhomogeneous coordinate on $\mathbb{CP}^1$ is given by the ratio:
\be
Z = \tan\frac{\theta}{2}e^{i\xi_{21}} \,.
\ee
The Fubini-Study metric is
\begin{align}
ds^2_{FS} &= \frac{dZd\bar{Z}}{(1+|Z|^2)^2}
=\frac{1}{4}\left(d\theta^2+\sin^2\theta d\xi_{21}^2\right)\,.
\end{align}
It coincides with the round metric on the two sphere. The connection
one-form associated to the projective space is given by
\be
A_{FS} = -\frac{i}{2}\frac{Z d\bar{Z}-\bar{Z}dZ}{(1+|Z|^2)} = -\sin^2\frac{\theta}{2}\, d\xi_{21}\,.
\ee
Up to a Wilson line, we can also write
%C
%
\be
A_{FS} = \frac{1}{2}\cos\theta\, d\phi \,.
\ee
The four dimensional asymptotically linear dilaton metric we use is \cite{HK2}
\be
ds^2 = \frac{g_2(Y)}{2}dY^2 + \frac{1}{2g_2(Y)}(d\psi + \cos\theta d\phi)^2 + \frac{Y}{2}(d\theta^2+\sin^2\theta d\phi^2) \,.  
\ee
The change in Wilson line above, is equivalent to a redefinition of the angle $\psi$.

\subsubsection{Asymptotic linear dilaton fibration over $\mathbb{CP}^2$}

In the six-dimensional asymptotic linear dilaton background, we make the choice of coordinates:
\begin{align}
z_i &= \mu_i\, e^{i\xi_i} \quad i=1,2,3\quad \text{with}\cr 
\mu_1&= \cos\chi\quad \mu_2 = \sin\chi\, \cos\frac{\theta}{2}\quad \text{and}\quad\mu_3 = \sin\chi\sin\frac{\theta}{2} \,.
\end{align}
We choose the following inhomogeneous coordinates on $\mathbb{CP}^2$ \cite{Gibbons:1978zy}:
\begin{align}
Z_1 &= \tan\chi  \cos\frac{\theta}{2} e^{i\frac{\varphi + \phi}{2}} \qquad\text{and}\qquad
Z_2 = \tan\chi  \sin\frac{\theta}{2} e^{i\frac{\varphi - \phi}{2}} \,.
\end{align}
The connection one-form takes the  form
\be
A_{FS}= -\frac{1}{2}\sin^2\chi\, (d\varphi + \cos\theta d\phi) \,.
\ee
The Fubini-Study metric 
is:
\be
ds^2_{FS} = d\chi^2 + \frac{1}{4}\sin^2\chi \cos^2\chi (d\varphi +\cos\theta d\phi)^2 + \frac{1}{4}\sin^2\chi (d\theta^2+ \sin^2\theta d\phi^2) \,.
\ee
A vierbein for $\mathbb{CP}^2$ is therefore given by
\begin{align}
e^1 &= d\chi\qquad e^2 = \frac{1}{2}\sin\chi \cos\chi (d\varphi + \cos\theta d\phi) \cr
e^3 &= \frac{\sin\chi}{2} d\theta \qquad e^4 = \frac{1}{2} \sin\chi \sin\theta d\phi \,.
\end{align}
The six-dimensional asymptotically linear dilaton
 spaces are fibrations of the form \eqref{generalALD} with $N=3$:
\begin{align}
ds^2 &= \frac{g_3(Y)}{2}dY^2 + \frac{2}{ g_3(Y)}\left(\frac{d\psi}{3} + A_{FS}\right)^2 + 2Y ds^2_{FS}\,,  
\end{align}

\section{Euclidean Taub-NUT Backgrounds}\label{ETN}

In this appendix we perform the fermionic zero mode counting in the
Euclidean Taub-NUT background with self-dual gauge field. The
calculation of zero modes for this background has been carried out in
\cite{Pope}. We will revisit the calculation and arrive at the same
result in a mildly more efficient manner. We moreover extend the calculation
to the case of the heterotic Taub-NUT dyon.
\subsection{Euclidean Taub-NUT with self-dual gauge field}

The metric of the Euclidean Taub-NUT space is given by 
\be
ds^2 = \frac{r+M}{r-M}dr^2 + 4M^2\frac{r-M}{r+M}(d\psi+\cos\theta d\phi)^2+(r^2-M^2)(d\theta^2+\sin^2\theta d\phi^2),
\ee
where $M$ is a parameter with the dimension of mass.  This can be put
in the form \eqref{general4dmetric} by using the radial variable
$x=\frac{r-M}{2M}$:
\be
\frac{ds^2}{4M^2} = \frac{(x+1)}{x}dx^2+\frac{x}{x+1}(d\psi + \cos\theta d\phi)^2 
+ x(x+1)(d\theta^2+\sin^2\theta d\phi^2)\,.
\ee
We can set $f(x) = x(x+1)$, $g(x)=\frac{x+1}{x}$ and
$M^2=\frac{1}{8}$. For easier comparison with the results of \cite{Pope}, we will keep
$M$ arbitrary in what follows. We have a self-dual gauge field
given by
\be
A = -P \frac{x}{x+1} (d\psi+\cos\theta d\phi)\,.
\ee
We note that this gauge field is proportional to the dual of the
Killing vector $\partial_\psi$ and thus fits the analysis of section
\ref{sqm}. 
The massless Dirac equation can be solved as in the examples discussed in the main text. The relevant linear differential operators that act on the positive chirality spinor components are given by
\begin{align}
-L_3&=x\frac{d}{dx} +ePx-(1+x)n +\frac{(1+2x)}{1+x}\,,&\qquad
L_4=x\frac{d}{dx} -ePx+(1+x)n +\frac{(1+2x)}{1+x}\cr
%C
N_3&=\frac{d}{d\theta}-m\csc\theta+\left(\frac{1}{2}+n\right)\cot\theta\,,&\qquad
N_4=\frac{d}{d\theta}+m\csc\theta+\left(\frac{1}{2}-n\right)\cot\theta\,.
\end{align}
We solve the equations $L_3(x) \cdot G_3(x) = 0$ and $N_4(\theta)
\cdot S_3(\theta)=0$ (and similarly for the other components) in order
to find the normalizable zero modes. This is more efficient than turning to the second order
differential equation as in \cite{Pope}. The solutions take the form
\be
\Psi_+ = e^{i n\psi+i m \phi}\begin{pmatrix}
e^{x(-eP+n)}x^{-\frac{1}{2}+n}(1+x)^{-\frac{1}{2}}(\sin\theta)^{-\frac{1}{2}+n}(\tan\frac{\theta}{2})^{-m}\cr
e^{x(eP-n)}x^{-\frac{1}{2}-n}(1+x)^{-\frac{1}{2}}(\sin\theta)^{-\frac{1}{2}-n}(\tan\frac{\theta}{2})^{m}
\end{pmatrix}
\ee
The wave functions agree with \cite{Pope}. For positive $eP$, we therefore
find normalizable solutions for all integers $l$ which satisfy the
conditions
\be
2|n| = 2l+1 \le 2[eP] \,.
\ee
There are no
normalizable solutions with the opposite chirality.
The counting function for positive $eP$ can be summarized as:
\be
Z_{ETN} = \sum_{2 n=1,2, \dots}^{2 {[} e P {]} } 2n\, y^n\,.
\ee
{From} subsection \ref{physics} (and the remark on radial asymptotics in appendix \ref{lineareqns}), it should be clear that 
the physics underlying the counting function can be summarized by saying that we 
restrict to the lowest Landau level on a
$\mathbb{CP}^1$ slice near the tip, that the degeneracy of the level
 is determined by its spin and 
that the bound state momentum is restricted by radial
asymptotic normalizability

\subsection{The heterotic Euclidean Taub-NUT dyon}
In this subsection, we review the construction of the heterotic Euclidean Taub-NUT dyon,
and calculate the Dirac index in this background.
\subsubsection{The background}
The starting point for the construction of the  heterotic Taub-NUT dyon
\cite{Johnson:1994ek} is the Taub-NUT metric and a constant
dilaton which gives rise to a solution of low-energy heterotic string theory:
\be
ds^2=\frac{ dr^2}{f_1(r)} + 4M^2f_1(r)(d\psi+\cos\theta d\phi)^2 + (r^2-M^2)(d\theta^2+\sin^2\theta d\phi^2)
\ee
where
\be
f_1(r) = \frac{r-M}{r+M}.
\ee
As in the bulk of the paper, we use the $SO(2)$ duality symmetry
of the space of supergravity solutions to construct a dyonic
generalization of the Taub-NUT metric \cite{Johnson:1994ek}. The dyonic Taub-NUT metric
is given by
\be 
ds^2 = \frac{dr^2}{f_1(r)}
+ 4M^2 \b^2 \frac{f_1(r)}{f_2(r)^2}( d\psi +\cos\theta
d\phi)^2 + (r^2-M^2)(d\theta^2+\sin^2\theta d\phi^2)\,, 
\ee 
where the function $f_2$ is
\be
f_2 = \frac{r+(2\b-1) M}{r+M}.
\ee
%
%When we set $\b =1$, we get $f_2=1$ and we recover the Euclidean Taub-NUT metric. 
We prefer to work in with the coordinate $x = \frac{r- M}{ 2M}$, in
terms of which the metric takes the form
\be
\frac{ds^2}{4M^2} = \frac{1+x}{x}dx^2 + \frac{x(1+x)}{(x+\beta)^2}
\b^2 (d\psi+\cos\theta d\phi)^2 + x(x+1)(d\theta^2+\sin^2\theta d\phi^2)\,.
\ee
The duality rotation also induces other background fields. These are given by 
\begin{align}\label{hetrotbkgnd}
B_{\psi \phi} &= 4M^2\frac{(1-\beta)x}{x+\beta} \cos \theta \qquad e^{- 2\Phi} = f_2(x) = \frac{x+\beta}{x+1} \cr
A_\psi &= 8M^2\frac{\sqrt{\beta(1-\b)}}{x+\b}  \qquad
A_\phi = -8M^2\frac{ \sqrt{\beta(1-\b)} x}{x+\b} \cos \theta \,.
\end{align}
The gauge field components can be combined to write
\be
A = 8M^2\sqrt{\frac{1-\b}{\b}}\left(d\psi -\frac{x}{x+\b}(d\psi +\b \cos\theta d\phi) \right)\,.
\ee
We will neglect the first (pure gauge) term in the analysis of the
Dirac equation. In order for the metric to be non-singular near the
tip of the space, we impose the periodicity condition $\psi \equiv
\psi + 4 \pi$. This turns the duality rotated background
into a genuinely new solution. Indeed, a continuous
duality transformation, supplemented with new discrete identifications in
the rotated solution, leads to a physically distinct background.

\subsubsection{Dirac index for a charged spinor on the Taub-NUT dyon}
In this subsection, we calculate the Dirac index for a charged spinor in the 
Euclidean Taub-NUT dyon background. By choosing the orthonormal real vierbeins
\begin{align}
e^1 &= \frac{2M dx}{\sqrt{f_1(x)}}\qquad e^2= 2M \b \frac{\sqrt{f_1(x)}}{f_2(x)}( d\psi+\cos\theta d\phi)\cr
e^3 &=2M\sqrt{x(x+1)}d\theta\qquad e^4 = 2M\sqrt{x(x+1)} \sin\theta d\phi\,,
\end{align}
one can check that %, like for the ordinary Taub-NUT case, 
 the field strength can be written as
\begin{align}
F
&=\frac{2\sqrt{\b(1-\b)}}{(x+1)(x+\b)}(e^3\wedge e^4-e^1\wedge e^2)\,.
\end{align}
The electromagnetic field $F$ carries electric and magnetic charge and behaves
like
$F_{x\psi} \approx Q/x^2$ and $F_{\theta\phi} \approx Q
\sin\theta$ at $x \rightarrow \infty$, where
\be
Q = 8M^2\sqrt{\b(1-\b)}\,. 
\ee
A gauge transformation brings 
the gauge field in equation \eqref{hetrotbkgnd}
into the form
\begin{align}
A %&=-8M^2\sqrt{(1-\b)\b}\frac{x}{x+\b}( d\psi+\cos\theta d\phi)\cr
&=-\frac{Qx}{x+\b}(d\psi+\cos\theta d\phi)\,.
\end{align}
The component Dirac equations can be solved as before 
by  separating variables. The positive chirality wave functions are given by
\be
\Psi_+= e^{in\psi + i m \phi}
\frac{ (x+\b)^{\frac{1}{4}}}{(1+x)^{\frac{3}{4}}}\begin{pmatrix}
e^{\frac{x}{\b}
\left(eQ +n \right)}
(\sin\theta)^{-\frac{1}{2}+n} (\tan\frac{\theta}{2})^{-m}\cr
%%%
e^{-\frac{x}{\b}\left(eQ  +n\right)} 
(\sin\theta)^{-\frac{1}{2}-n} (\tan\frac{\theta}{2})^{m}
\end{pmatrix}\,.
\ee
The index generating function now equals:
\begin{align}
Z_{ETN}^{het} &=
\sum_{2 n=1,2, \dots}^{ {[}  - 2 e Q {]} } 2n \, y^n
\,.
\end{align}
for 
$eQ<0$.

\section{Conventions for Spinors}
\label{spinors}

In two dimensions, we use the basis of gamma-matrices
\begin{align}
\gamma^1 = \sigma_1 = \begin{pmatrix}0 &1 \\ 1 & 0\end{pmatrix} \qquad
\gamma^2 = \sigma^2 = \begin{pmatrix}0 &-i \\ i & 0\end{pmatrix} \qquad
\gamma^3 = i\gamma^1\gamma^2 = -\sigma_3=\begin{pmatrix}-1 &0 \\ 0 & 1\end{pmatrix}\,.
\end{align}
In four dimensions, we choose a Weyl basis for the gamma matrices:
\begin{align}
\gamma^k = 
\begin{pmatrix}
0 &\sigma^k  \cr
-\sigma^k & 0 \cr
\end{pmatrix} \qquad
\gamma^4 =\begin{pmatrix}  0 & \mathbb{I}_2 \cr 
\mathbb{I}_2 & 0\end{pmatrix} \qquad
\gamma^5  = \gamma^1 \gamma^2 \gamma^3 \gamma^4 
=\begin{pmatrix} 
-\mathbb{I}_2 & 0 \cr 
0 & \mathbb{I}_2  \end{pmatrix} \,.
\end{align}
Our ansatz for the positive chirality and negative chirality spinors in four dimensions is given by:
\be
\Psi=\begin{pmatrix}
\Psi_- \cr
\Psi_+
\end{pmatrix}
\ee
where the two component Weyl spinors are given by
\be\label{4dweyl}
\Psi_- = e^{i n \psi+i m\phi}
\begin{pmatrix}
G_1(Y) S_1(\theta)\cr
G_2(Y) S_2(\theta)
\end{pmatrix}
\qquad \text{and}\qquad
\Psi_+ = e^{i n \psi+i m\phi}
\begin{pmatrix}
G_3(Y) S_4(\theta)\cr
G_3(Y) S_4(\theta)
\end{pmatrix}\,.
\ee
Finally, in six dimensions we construct the gamma matrices by tensoring the four
dimensional gamma matrices with two-dimensional  Pauli matrices as follows:
\begin{align}
\Gamma^1 &= \gamma^1\otimes \sigma_3\qquad
\Gamma^2 = \gamma^2 \otimes \sigma_3\qquad
\Gamma^3 = \gamma^3\otimes \sigma_3\cr
\Gamma^4 &= \gamma^4 \otimes \sigma_3\qquad
\Gamma^5 = \mathbb{I}_4 \otimes \sigma_1\qquad
\Gamma^6 = \mathbb{I}_4 \otimes \sigma_2 \,.
\end{align}
The matrix $\Gamma^7$ is determined by $\sqrt{-1}$ times the product of all the gamma matrices. 
After permuting the third and seventh and first and fifth rows and columns,
the  matrix $\Gamma^7$ takes the canonical form:
\be
 \Gamma^7 = \begin{pmatrix}
-\mathbb{I}_7 & 0 \cr
0 & \mathbb{I}_7
\end{pmatrix}
\ee
In this Weyl basis, the Dirac spinor is of the form
\be
\Psi=\begin{pmatrix}
\Psi_-\cr
\Psi_+
\end{pmatrix}
\ee
where the four component Weyl spinors in six dimensions are 
\be\label{Weylspinor6d}
\Psi_{-} = e^{in \psi + ip \varphi+i m \phi}
\begin{pmatrix}
G_1(Y) T_1(\chi) S_1(\theta)\\
G_2(Y) T_2(\chi) S_2(\theta)\\
G_3(Y) T_3(\chi) S_3(\theta)\\
G_4(Y) T_4(\chi) S_4(\theta)
\end{pmatrix}\quad\text{and}\quad
\Psi_{+} = e^{in \psi + ip \varphi+i m \phi}
\begin{pmatrix}
G_5(Y) T_5(\chi) S_5(\theta)\\
G_6(Y) T_6(\chi) S_6(\theta)\\
G_7(Y) T_7(\chi) S_7(\theta)\\
G_8(Y) T_8(\chi) S_8(\theta)
\nonumber
\end{pmatrix}\,.
\ee

\end{appendix}

\end{document}